\documentclass{jpp}
\usepackage{graphicx}

\usepackage[utf8]{inputenc}
\usepackage[T1]{fontenc}
\usepackage{natbib}
\usepackage{amsmath}
\usepackage{etoolbox}
\usepackage{url}
\patchcmd{\subequations}{\alph{equation}}{\textit{\alph{equation}}}{}{}

\usepackage{cleveref}
\usepackage{bm}
\usepackage{mathtools}

\shorttitle{Comprehensive full-f drift-kinetic and delta-f gyrokinetic simulations}
\shortauthor{J. E. Mencke, P. Ricci, and L. Da Silva}

\title{Comprehensive full-f drift-kinetic and delta-f gyrokinetic simulations of a linear plasma device based on the gyro-moment approach}

\author{J. E. Mencke\aff{1}
  \corresp{\email{jacob.mencke@epfl.ch}},
  \and P. Ricci\aff{1}
  \and L. Da Silva\aff{1}}

\affiliation{\aff{1}Ecole Polytechnique F\'ed\'erale de Lausanne (EPFL), Swiss Plasma Center, CH-1015 Lausanne, Switzerland}

\usepackage{physics}

\newcommand{\R}{\bm R}

\newcommand{\Bpar}{B_{\|}^{*}}

\newcommand{\A}{\bm A}

\newcommand{\N}{\mathcal{N}}

\newcommand{\C}{ \mathcal{C}}

\renewcommand{\|}{\parallel}

\newcommand{\projpj}[2]{ \norm{#2}^{#1}}

\renewcommand{\poissonbracket}[2]{  \left[#1, #2 \right]}
\newcommand{\gyroavg}[1]{\left\langle #1\right\rangle_{\R}}
\newcommand{\gyroavgadj}[1]{\left\langle #1\right\rangle_{\bm x}^{\dagger}}
\renewcommand{\eqref}[1]{Eq. (\ref{#1})}
\newcommand{\appref}[1]{App. \ref{#1}}
\newcommand{\secref}[1]{Sec. \ref{#1}}
\newcommand{\figref}[1]{Fig. \ref{#1}}

\begin{document}

\maketitle

\begin{abstract}
    First of a kind comprehensive full-f drift-kinetic (DK) and $\delta$-f gyrokinetic (GK) turbulent simulations are carried out in a linear plasma device. We self-consistently derive an electrostatic model including large-scale slowly-varying DK-ordered fields coupled to small-scale rapidly-fluctuating GK-ordered fields. By relying on the critical balance ordering, we show that the electrons are described by a drift-reduced Braginskii model while we rely on a Hermite-Laguerre spectral expansion for describing both the DK and GK parts of the ion distribution function. Global simulations are carried out using the parameters of the linear device LAPD, showing that the DK part of the ion distribution function is approximately a bi-Maxwellian. Fast spectral convergence both for the DK and GK Hermite-Laguerre expansion coefficients is observed, and that the GK fields do not affect the DK fields at the physical LAPD collisionality. Only when the collisionality is reduced and the source term is amplified for the GK fluctuations, an amplification of small-scale turbulent structures is observed. The findings are supported by linear results that show that the simulations are dominated by turbulent fluctuations that are Kelvin-Helmholz driven. Additionally, a GK Kelvin-Helmholz-like mode is observed in the low-GK-collisionality regime which can non-linearly drive small-scale structures.
\end{abstract}

\section{Introduction}\label{sec:introduction}
Modeling the boundary region of magnetically confined fusion plasmas remains a challenge for the fusion community. Fluctuations at all scales need to be considered, as well as kinetic effects and collisional processes. Developing a model for simulating the plasma boundary requires overcoming the limitations of existing models. Full-f fluid codes focus on the long-wavelength limit \cite{Paruta2018,Giacomin2022}, and their validity is limited to the highly collisional regime. Gyrofluid codes apply Pad\'e approximation to relax the long-wavelength approximation \cite{held2022beyond,held2020pade}. Additionally, they use kinetic closures to avoid considering the highly collisional limit, although the success of these closures is highly case dependent \cite{Beer1996,wiesenberger2022long,wiesenberger2023effects}. On the other hand, gyrokinetic (GK) $\delta$-f models, which offer a good description of the core of fusion plasmas \cite{Pan2016,jenko2000electron}, present the challenge of including collisions at the level of relevance for the boundary. Furthermore, their application to the boundary is questionable due to large-scale fluctuations present in this region. Full-f GK codes often focus on the long-wavelength limit \cite{Chang2017,Dorf2020,Hakim2020,Michels2022,Ulbl2023} and thus compromise physical fidelity at the cost of computational performance. 

To overcome the limitations of these approaches and accurately describe the boundary, \cite{qin2006general,hahm2009fully,Frei2020}, propose a model that couples a DK and GK ordering. Here, we leverage the work of \cite{Frei2020} and present a first implementation of a model that couples DK and GK orderings. More precisely, this work relies on splitting the distribution functions into a drift-kinetic (DK) large-scale slowly-varying part and a GK small-scale rapidly-varying part. The DK fields are allowed to be large in amplitude but only vary on long scale, employing a long-wavelength limit, while the GK fields are small in amplitude but allowed to vary on all scales.

The model is based on single particle equations of motion derived from first principles starting from the particle one-form. From the single particle equations of motion, the gyro-averaged Boltzmann equation is derived, split into a GK part and a DK part, and projected onto a Hermite-Laguerre basis giving a DK and a GK hierarchy equation for the projection coefficients, denoted as gyromoments. The Hermite-Laguerre projection allows for an efficient implementation of the plasma dynamics, including multiple different collision operators. Finally, we split the electrostatic potential into a DK part and a GK part, similarly to the distribution functions, and we derive a DK and GK Poisson equation from first-principles by using a variational principle. 

Previous $\delta$-f GK gyromoment simulations based on the Hermite-Laguerre projection are shown to efficiently reduce the computational complexity of solving the GK system of equations compared to continuum $\delta$-f GK simulations \cite{Hoffmann2023,Hoffmann2023b,Mandell2022}. Similarly, full-f gyromoment simulations were performed in a linear plasma device, within a DK ordering \cite{frei2023fullf,mencke2025extended}. This work combines the advantages of the two previously developed models allowing for order-one DK fluctuations with small wavenumbers and, at the same time, small-amplitude GK fluctuations with arbitrary wavenumbers. It then enables the study of instabilities on all scales while still removing the single-particle gyromotion from the description.


Although our approach ultimately is intended for the boundary of magnetically confined fusion devices, we consider the simpler case of a linear plasma device in this work. A linear plasma device offers an ideal testbed for new global turbulent plasma simulations. The absence of drifts associated with the curvature of the magnetic field, together with the smaller size of the simulations compared to toroidal devices reduces the complexity and cost of the simulations and allows for a deeper understanding of the physics at play \cite{Friedman2012}. We note that, despite their simple geometries, previous simulations of linear plasma devices show complex turbulent behavior \cite{Shi2017,Pan2016,Pan2018,Popovich2010b,Rogers2010,frei2023fullf,mencke2025extended}. Finally, the perpendicular incidence of the magnetic field with the walls at the end of the plasma column simplify the boundary conditions of the simulations.

This paper is structured as follows. In section \secref{sec:oredering} we present the ordering used as well as the equations describing the evolution of the distribution functions. The DK model for the ions and electrons is derived in \secref{sec:DK_sys} from a kinetic description, and similarly, the GK model is derived in \secref{sec:GK_sys}. We close the system in \secref{sec:field_equations}, by deriving the DK and GK Poisson equations used to solve for the two electrostatic potentials. The numerical implementation of the physical model is described in \secref{sec:Num_impl}. In \secref{sec:sim_res} we report the results of the comprehensive turbulent simulations comparing it with a full-f DK simulation where the GK perturbations are suppressed. Section \ref{sec:linear} provides an analysis of the findings in \secref{sec:sim_res} by presenting linear investigations of the turbulence-driving instabilities. Finally, the conclusions follow in \secref{sec:concl}.
 
\section{Ordering assumptions and equations of motion}\label{sec:oredering}
We focus on the case of a fully-ionized plasma with one single-ionized ion species such that the plasma is composed of species $a=i,e$ with $e$ denoting the electrons and $i$ the only ion species with charge $q_i=e$.
We split the electrostatic potential
\begin{equation}
    \phi\left(\mathbf{x}\right)= \phi_{DK}\left(\mathbf{x}\right)+\phi_{GK}\left(\mathbf{x}\right),\label{eq:split_phi_GK}
\end{equation}
where $\mathbf{x}$ is the particle position and we impose that the GK part of the electrostatic potential has a small amplitude compared to the DK part, $\phi_{GK}/\phi_{DK}\sim \epsilon_{\delta}<1$. We order the DK and GK quantities such that they vary on different length scales. More precisely, we order the perpendicular gradient lengths, $k_{\perp DK}\sim \left|\nabla_{\perp} \ln F_{aDK}\right|\sim \left|\nabla_{\perp} \ln \phi_{DK} \right|\sim \epsilon_{\perp}/\rho_s$ with $\epsilon_{\perp}<1$ and $k_{\perp GK}\sim \left|\nabla_{\perp} \ln \phi_{GK} \right|\sim 1/\rho_s$, being $\rho_s=c_s/\Omega_i$ the ion sound Larmor radius with $c_s=\sqrt{T_e/m_i}$ the sound speed, $T_e$ the electron temperature, and $\Omega_i=q_i B/m_i$ the ion gyro frequency, with $B$ the magnetic field strength. Furthermore, we order the parallel length scale as $k_{\|}\sim \epsilon k_{\perp DK}$ and the frequency as $\omega\sim \Omega_i \epsilon\epsilon_{\perp}$. This ordering imposes that $\phi_{DK}$ gives a contribution to the $\bm E \times \bm B$ velocity on the order of $\epsilon_{\perp}/\epsilon_{\delta}$ with respect to that of $\phi_{GK}$ \cite{Frei2020}. By further ordering $\epsilon_{\perp}\sim \epsilon_{\delta}$, we impose that the $\bm E \times \bm B$ drift from the DK electrostatic potential contributes to the drifts, comparably to the GK counterpart. 

Based on this ordering, we derive the single-particle dynamics in \secref{subsec:sing-par}. The behavior of the full distribution function is described in \secref{subsec:col_beh} using the Boltzmann equation. Finally, the evolution of the DK and GK parts of the distribution function, is derived from the Boltzmann equation in \secref{subsec:DK_boltz}.

\subsection{Single-particle dynamics}\label{subsec:sing-par}
We assume that the collisionless particle dynamics be governed by the gyrocenter single-particle Lagrangian, $\Gamma_A$. By using a Lie approach and including all terms up to second order in the GK and DK parts of the $\bm E \times \bm B$ drift (i.e., to order $c_s^2 \epsilon_{\perp }^2$ and $c_s^2 \epsilon_{\delta}^2$), while assuming a straight and constant magnetic field, we have \cite{Frei2020}
\begin{align} \label{eq:Gammai}
\Gamma_a(\bm R, \mu, v_\parallel,t)= q_a \bm A_a^* \cdot \dot \R - \frac{\mu B}{\Omega_i} \dot \theta - m_a v_\parallel^2 / 2 - q_a \Phi_a^*+\mathcal{O}\left(\frac{m_a}{m_i}T_a\epsilon_{\perp}^3,\frac{m_a}{m_i}T_a\epsilon_{\delta}^3\right),
\end{align}
where $v_{\|}=\mathbf{v}\cdot\mathbf{b}$ is the velocity component parallel to the magnetic field, $\mathbf{b}$ being the unit vector in the direction of the magnetic field, $\mu = m_a v_{\perp}^2/B$ is the adiabatic invariant, $m_a$ is the mass of species $a$, and $v_{\perp}=\left|\mathbf{v}-\mathbf{b}v_{\|}\right|$ is the velocity component perpendicular to the magnetic field. We also define the fields $q_a \Phi_a^*  =q_a \phi_{DK} + \mu B+q_a \Phi_E+q_a\gyroavg{\psi} $ and $q_a \bm A_a^* = q_a \A +
 m_a v_\parallel \bm b+m_a \bm u_{E}$, where $q_a\Phi_E=m_a\left|\bm u_E\right|^2/2+\mu B \nabla_{\perp}^2\phi_{DK} /\left(2 B \Omega_a \right)$ and $\bm u_E=-\nabla \phi_{DK} \times \bm B/B^2$ is the $\bm E \times \bm B $ velocity for the DK potential. The effective GK gyro-averaged potential is defined as
 \begin{equation}
 \gyroavg{\psi} =\gyroavg{\phi_{GK}}+\frac{q_a^2}{2m_a\Omega_a}\partial_{\mu}\left(\gyroavg{\phi_{GK}}^2-\gyroavg{\phi_{GK}^2}\right)+\frac{q_a}{2m_a \Omega_a^2}\gyroavg{\left(\mathbf{b}\times\nabla \overline{\widetilde{\phi_{GK}}}\right)\cdot\nabla \widetilde{\phi_{GK}}},\label{eq:gyro_avg_psi}
 \end{equation}
 where the gyroaverage operator is introduced,
 \begin{subequations}
 \begin{align}
     \gyroavg{f}&=\frac{1}{2 \pi} \int_0^{2 \pi} \mathrm{d} \theta \int \mathrm{d} \boldsymbol{x} \delta({\boldsymbol{R}}+{\rho}-\boldsymbol{x}) f\left(\boldsymbol{x}, {\mu}, {v}_{\|}, t\right)\\
     &=\frac{1}{2 \pi} \int_0^{2 \pi} \mathrm{d} \theta \sum_{n=0}^{\infty} \frac{\left(\boldsymbol{\rho}_a\cdot \nabla\right)^n}{n!}f\left(\boldsymbol{R}, {\mu}, {v}_{\|}, t\right)\\
     &=\sum_{n=0}^{\infty}\left(\frac{2\mu}{q_a\Omega_a}\right)^{n}\frac{\nabla_{\perp}^{2n}}{2^n n!n!}f\left(\boldsymbol{R}, {\mu}, {v}_{\|}, t\right),
 \end{align}
 \end{subequations}
 with $\delta\left(\cdot\right)$ the Dirac delta function and $\boldsymbol{\rho}_a=\mathbf{a}v_{\perp}/\Omega_a$, the Larmor radius, being $\mathbf{a}$ a unit vector perpendicular to both the magnetic field and the gyromotion of the particle \cite{Frei2020}. Finally, we define

 \begin{subequations}
 \begin{align}
     \widetilde{\phi_{GK}}&=\phi_{GK}-\gyroavg{\phi_{GK}},
\intertext{and}
     \overline{\widetilde{\phi_{GK}}}&= \int^\theta d\theta^\prime \widetilde{\phi_{GK}}\left(\boldsymbol{x}\left(\bm R,\theta^\prime\right), {\mu}, {v}_{\|}, t\right),
 \end{align}
 \end{subequations}
 in \eqref{eq:gyro_avg_psi}.

\subsection{Collective behavior}\label{subsec:col_beh}

 The collective behavior of the distribution function is governed by the gyroaveraged Boltzmann equation \cite{Frei2020},
\begin{equation}
    \partial_t F_a+\nabla\cdot\left(\dot{\R}F_a\right)+\dot{v}_{\|}\partial_{v_{\|}} F_a =  \gyroavg{S_{a}}+ \sum_{b=i,e}\gyroavg{C_{ab}},\label{eq:Boltzmann_a}
\end{equation}
where the sources for species $a$, $S_a$, are introduced along with the collision operator between species $a$ and $b$, $C_{ab}$, the parallel acceleration, $\dot{v_{\|}}$, and gyrocenter drift, $\dot{\R}$. 

 The Euler-Lagrange equations of the ion Lagrangian, \eqref{eq:Gammai}, give
\begin{subequations}
\begin{align}
v_{\| } &=\bm b \cdot\dot{\R},\\
\dot{\mu}&=0,\\
-q_a \bm B^*\times \dot{\R}-m_a \dot{v}_{\|}\bm b&=q_a\nabla \Phi^*_a+q_a\partial_t \bm A^*,\label{eq:Euler_Lagrange_R}
 \end{align}
 \end{subequations}
 introducing the effective magnetic field, $\mathbf{B}^*=\nabla \times \mathbf{A}^*$. Now, projecting \eqref{eq:Euler_Lagrange_R} on the $\mathbf{B}^*$ direction and crossing with $\bm b$, we obtain respectively
 \begin{subequations}
 \label{eq:dot_v_and_R}
 \begin{align}
\dot{v}_{\|} \Bpar&=-\frac{q_a}{m_a} \bm B^*\cdot \left(\nabla \Phi^*_a+\partial_t\bm A^*\right),\label{eq:vpardot}\\
\dot{\R}\Bpar&=\bm B^* v_{\|}+\bm b\times \left(\nabla \Phi^*_a+\partial_t \bm A^*\right),\label{eq:Rdot}
 \end{align}
 \end{subequations}
where $\Bpar=\mathbf{B}^*\cdot\bm b$. Equations (\ref{eq:vpardot}) and (\ref{eq:Rdot}) are used to evolve the distribution functions through the Boltzmann equation.

Considering the lowest order in $m_e/m_i$ in \eqref{eq:Gammai}, the electron one-form becomes
\begin{equation}
    \Gamma_e(\bm R, \mu, v_\parallel,t)= q_e \bm A \cdot \dot \R - \frac{\mu B}{\Omega_e} \dot \theta - m_e v_\parallel^2 / 2 - q_e \left(\phi_{DK}+\phi_{GK}\right)+\mathcal{O}\left(\frac{m_e}{m_i}T_a\right),\label{eq:gamma_e}
\end{equation}
with the equations of motion identical to that of the ions in \eqref{eq:dot_v_and_R}, with $\Bpar=B$, $m_a=m_e$, $\Phi_e^*=\phi_{DK}+\phi_{GK}-\frac{\mu B}{e}$, $\bm B^*=\bm B$, and $\partial_t\bm A^*=0$.


\subsection{DK and GK Boltzmann equations}\label{subsec:DK_boltz}

Given the equations of motion in \eqref{eq:dot_v_and_R}, \eqref{eq:Boltzmann_a} can, in principle, be solved with appropriate numerical methods once the sources, $S_a$, and collision operators, $C_{ab}$, are defined. However, we rely, in the following, on the splitting of the distribution function in a similar way as $\phi$ in \eqref{eq:split_phi_GK}. That is
\begin{equation}
    F_a\left(\mathbf{R},v_{\|}, \mu\right)= F_{aDK}\left(\mathbf{R},v_{\|}, \mu\right)+F_{aGK}\left(\mathbf{R},v_{\|}, \mu\right).\label{eq:Fa_split}
\end{equation}
Analogously to the electrostatic potential, we impose, $F_{aGK}\sim \epsilon_{\delta }F_{aDK}$, $\left|\nabla_\perp \ln F_{aDK}\right|\sim k_{\perp DK}$, and $\left|\nabla_\perp \ln F_{aGK}\right|\sim k_{\perp GK}\sim 1/\rho_s$.

To separate the DK and GK part of the distribution function, we average \eqref{eq:Boltzmann_a} over small scales, defining the averaging operator such that for any DK-ordered field, $g_{aDK}$,
\begin{subequations}
    \begin{align}
        \left\langle g_{aDK}\right\rangle_{DK}&=g_{aDK},
        \intertext{and for the GK fields}
        \left\langle F_{aGK}\right\rangle_{DK}&=0,\\
        \left\langle\nabla_{\perp} \gyroavg{\phi_{GK}}\right\rangle_{DK}&=\nabla_{\perp}\phi_{GK},\\
        \left\langle\nabla_{\|} \gyroavg{\phi_{GK}}\right\rangle_{DK}&=0,
    \end{align}
\end{subequations}
with $\nabla_{\|}=\bm b \cdot \nabla$. The averaging operation removes all fluctuations on the Larmor-radius scale and we assume that only the perpendicular electric field has a large-scale component due to the assumption $\nabla_{\perp}\phi_{DK}\sim\nabla_{\perp}\phi_{GK}$. The leading order (in $\epsilon_\perp$) DK average of \eqref{eq:Boltzmann_a} is
\begin{subequations}
\begin{align}
    \partial_t F_{aDK}&+\nabla\cdot\left(\dot{\bm R}_{DK} F_{aDK}\right)+\dot{v}_{\| DK}\partial_{v_{\|}} F_{aDK} =  S_{aDK}+ \sum_{b=i,e}C_{ab DK}+\mathcal{O}\left(F_{a}\Omega_i\epsilon\epsilon_{\perp}^2\right),\label{eq:Boltzmann_DKa}
\intertext{where}
    \dot{\bm R}_{DK}&=\bm b v_{\|}+\frac{1}{B}\bm b\times\nabla\left(\phi_{DK}+\phi_{GK}\right),\quad \dot{v}_{\| DK}=\frac{q_a}{m_a}\nabla_{\|}\phi_{DK},\\
    S_{aDK}&=\left\langle S_a\right\rangle_{DK},\quad C_{abDK}=\left\langle C_{ab}\right\rangle_{DK}=C_{ab}\left(F_{aDK},F_{bDK}\right).
\end{align}
\end{subequations}

Finally, the GK Boltzmann equation is retrieved by subtracting \eqref{eq:Boltzmann_DKa} from \eqref{eq:Boltzmann_a}. That is,
\begin{subequations}
\begin{align}
    \partial_t F_{aGK}&+\nabla\cdot\left(\left(\dot{\bm R}-\dot{\bm R}_{DK}\right) F_{aDK}+\dot{\bm R} F_{GK}\right)+\dot{v}_{\| DK}\partial_{v_{\| }} F_{aGK}+ \dot{v}_{\| GK}\partial_{v_{\| }} F_{aDK}\nonumber\\
    &=  \gyroavg{S_{a}}-S_{aDK}+ \sum_{b=i,e}\left(\gyroavg{C_{ab}}-C_{ab DK}\right)+\mathcal{O}\left(F_{a}\Omega_i\epsilon\epsilon_{\perp}^2\epsilon_{\delta}\right),\label{eq:Boltzmann_GKa}
\intertext{with}
    \dot{v}_{\| GK}&=\frac{q_a}{m_a}\nabla_{\|}\gyroavg{\phi_{GK}}.
\end{align}
\end{subequations}

Eqs. (\ref{eq:Boltzmann_DKa}) and (\ref{eq:Boltzmann_GKa}) describe the evolution of the distribution functions given the electrostatic potentials, $\phi_{DK}$ and $\phi_{GK}$.
\section{Drift-kinetic governing equations}\label{sec:DK_sys}

This section presents the derivation of the part of the model that governs the dynamics of the distribution functions on the DK scale. We simplify the electron evolution described by the DK Boltzmann \eqref{eq:Boltzmann_DKa} using the ordering in \secref{subsec:Electron_DK}. Furthermore, also starting from \eqref{eq:Boltzmann_DKa}, we derive a gyromoment hierarchy equation for the ions in \secref{subsec:Ion_moment_DK} by expanding the ion distribution function on a Hermite-Laguerre basis.


\subsection{Electron drift-kinetic model}\label{subsec:Electron_DK}
For the electrons, we notice that the leading order term in \eqref{eq:Boltzmann_DKa} is order $F_{a}\Omega_i\epsilon\epsilon_{\perp}\sqrt{m_i/m_e}$. Expanding the electron distribution function in terms of square root of the mass ratio and keeping terms only to first order, $F_{eDK}=F_{eDK0}+F_{eDK1}$, with $F_{eDK1}\sim \epsilon_m F_{eDK0}$ and $\epsilon_m=\sqrt{m_e/m_i}$, the lowest order terms appearing in \eqref{eq:Boltzmann_DKa} are
\begin{subequations}
\begin{align}
    v_{\|}\nabla_{\|} F_{eDK0}&-\frac{e}{m_e}\nabla_{\|}\phi_{DK}\partial_{v_{\|}} F_{eDK0}=C_{ee}\left(F_{eDK0},F_{eDK0}\right)+C_{ei}\left(F_{eDK0},F_{iDK}\right),\label{eq:Boltzmanne0}
    \intertext{where $C_{ei}$, at leading order in $\epsilon_m$, is given by (\cite{braginskii1965transport})}
    C_{ei}\left(F_{eDK}, F_{iDK}\right)&=-\frac{L_{ei}N_{eDK}}{8\pi}\partial_{v_{\gamma}}\left(V_{\beta\gamma}\partial_{v_{\beta}}F_{eDK}\right) =\mathcal{L}_{ei}\left(F_{eDK}\right) \label{eq:coll_0_brag} , \\
\intertext{with}
    L_{ei}&=\frac{3\sqrt{\pi}}{4}\frac{\nu_{ei}}{N_{eDK}}v_{Th e}^3,\quad v_{Th e}=\sqrt{\frac{2 T_e}{m_e}},\\
    V_{\beta\gamma}&=\frac{1}{v^3}\left(v^2 \delta_{\beta \gamma}-v_{\beta}v_{\gamma}\right).
\end{align}
\end{subequations}
A solution of \eqref{eq:Boltzmanne0} is given by
\begin{subequations}
    \begin{align}
    F_{eDK0}&=N_{e0}\left(\mathbf{R},t\right)\frac{m_e^{3/2}}{\left(2\pi T_{e0}\left(\mathbf{R}_\perp,t\right)\right)^{3/2}}\exp\left(-\frac{m_ev^2}{2T_{e0}\left(\mathbf{R}_\perp,t\right)}\right),\\
    N_{e0}\left(\mathbf{R},t\right)&=N_{e\perp}\left(\mathbf{R}_\perp,t\right)\exp\left(-\frac{e\phi_{DK}}{T_{e0}\left(\mathbf{R}_\perp,t\right)}\right).\label{eq:Ne0_def}
\end{align}
\end{subequations}
In order to obtain an evolution equation for $N_{e0}$ and $T_{e0}$, we consider the next order in $\epsilon_m$ of the Boltzmann equation, \eqref{eq:Boltzmann_DKa}, that is
\begin{align}
    \partial_t F_{eDK0}&+\frac{1}{B}\poissonbracket{\phi_{DK}+\phi_{GK}}{F_{eDK0}}+v_{\|}\nabla_{\|} F_{eDK1}-\frac{e}{m_e}\nabla_{\|}\phi_{DK}\partial_{v_{\|}} F_{eDK1}\nonumber\\
    &=C_{ee}^{\left(1\right)}\left(F_{eDK1}\right)+\mathcal{L}_{ei}\left(F_{eDK1}-\frac{m_e v_{\|} J_{\|}}{N_{e0}T_{e0}} F_{e0}\right)+S_{eDK},\label{eq:Boltzmanne1}
\end{align}
with $\poissonbracket{f}{g}=\bm b\cdot\nabla f \times \nabla g$. The $F_{eDK1}$, that solves \eqref{eq:Boltzmanne1}, is 
\begin{subequations}
\begin{align}
    F_{eDK1}&=\left[\frac{N_{e1}}{N_{e0}}+\frac{P_{e1}}{N_{e0}T_{e0}}\left(\frac{m_ev^2}{2T_{e0}}-\frac{3}{2}\right)\right]F_{eDK0}+f_{sh}\label{eq:FeDK1_ansatz},\\
\intertext{where}
    f_{sh}&=\sum_{k=0}^{\infty}v_{\|} \frac{F_{eDK}}{N_{e0}} A_k\left(\bm R\right) L_{k}^{3/2}\left(m_ev^2/2T_{e0}\right).\label{eq:fsh_ansatz}\\
\intertext{Truncating $f_{sh}$ at $k=3$, we determine $A_k$, as described in \appref{app:Fe1_calc}, obtaining
}
    A_0&=2\frac{m_e}{T_{e0}}N_{e0}U_{\| e DK},\quad A_1 =1.266 C_{P1}+0.286\frac{m_e}{T_e} J_{\|},\\
    A_2&= -0.654 C_{P1}+0.055 \frac{m_e}{T_e} J_{\|},\quad A_3=0.0193 C_{P1}+0.0238 \frac{m_e}{T_e} J_{\|},
\intertext{being}
    C_{P1}&=\frac{1}{T_{e0}\nu_{ee}}\nabla_{\|}T_{e1},\quad \text{and}\quad T_{e1}= \frac{P_{e1}}{N_{e0}}.
\end{align}
\end{subequations}
By taking the moments of \eqref{eq:Boltzmanne1} with weight $1$ and $m_ev^2/2$, we determine an evolution equation for $N_{eDK}=N_{e0}+N_{e1}$ and $T_{eDK}=T_{e0}+T_{e1}$, respectively, that is
\begin{subequations}\label{eq:Eletron_system}
\begin{align}
    &\partial_t  N_{eDK}+\frac{1}{B}\poissonbracket{\phi_{DK}+\phi_{GK}}{ N_{eDK}}+\nabla_{\|}\left(U_{\| eDK}N_{eDK}\right)=S_N+\mathcal{O}\left(N_{e}\Omega_i\epsilon\epsilon_{\perp}^3\right)\label{eq:dt_Ne},\\
\intertext{and}
    &\frac{3}{2}\left(\partial_t  T_{eDK}+\frac{1}{B}\poissonbracket{\phi_{DK}+\phi_{GK}}{ T_{eDK}}+U_{\|  eDK}\nabla_{\|}T_{eDK}\right)\nonumber\\
    &+T_{eDK}\nabla_{\|} U_{\| eDK}-\frac{0.71 T_{eDK}}{N_{eDK}}\nabla_{\|}J_{\|}+\frac{1}{N_{eDK}}\nabla_{\|}q_{\| e r}=\frac{3}{2}S_{Te}+\mathcal{O}\left(T_{e}\Omega_i\epsilon\epsilon_{\perp}^3\right),\label{eq:dt_Te}\\
\intertext{being}
    q_{\| e r}&=-\frac{3.16N_{eDK}T_{eDK}}{m_e\nu_{ee}}\nabla_{\|}T_{eDK},
\end{align}
\end{subequations}
(see \appref{app:Fe1_calc} for the details of the derivation). We note that Eqs. (\ref{eq:dt_Ne}) and (\ref{eq:dt_Te}) correspond to the evolution of the density and electron temperature in the drift-reduced Braginskii model with the electrostatic potential expressed as the sum of the GK and DK electrostatic potentials, $\phi_{DK}+\phi_{GK}$ \cite{braginskii1965transport,Zeiler1997,FelixParraBrag}.

\subsection{Ion moment hierarchy}\label{subsec:Ion_moment_DK}
For the ions, instead of ordering the ion-ion collision frequency as $\nu_{ii}\ll \omega$ and recover the drift-reduced Braginskii through a similar calculation as in \secref{subsec:Electron_DK}, we assume $\nu_{ii}\sim \omega$, similar to \cite{Frei2020,frei2023fullf,mencke2025extended}, and expand the DK part of the distribution functions on a Hermite-Laguerre basis
\begin{subequations}
\label{eq:Hpj_expansions}
\begin{align}
    F_{iDK}\left(\mathbf{R},v_{\|}, \mu\right)&=\sum_{p,j=0}^{\infty}\N^{pj}_{iDK}\left(\R\right)\frac{H_{p}\left(s_{\| i}\right)L_{j}\left(x_i\right)}{\sqrt{2^{p}p!}} F_{Mi}\left(s_{\| i},x_i\right),\label{eq:DK_Npj_exp}
\intertext{where we introduce}
    s_{\| i}&=\frac{v_{\|}-U_{\| iDK}\left(\R\right)}{v_{Th i}},\quad x_i = \frac{\mu B}{T_{i0}},\quad v_{Th i}=\sqrt{\frac{2T_{i0}}{m_i}},\\
    F_{Mi}&=\frac{e^{-s_{\|i}^2-x_i}}{\left(\sqrt{\pi}v_{Thi}\right)^3},
\end{align}
\end{subequations}
and the background temperature of the ions, $T_{i0}$, along with the physicists' Hermite polynomials, $H_p$, and the Laguerre polynomials, $L_j$. These polynomials satisfy the following orthogonality relations,
\begin{subequations}
\begin{align}
    \int_{-\infty}^{\infty}dx H_{p}\left(x\right)H_{p^{\prime}}\left(x\right) e^{-x^2}&=2^{p}p! \delta^{p}_{p^{\prime}},\\
    \int_{0}^{\infty}dx L_{j}\left(x\right)L_{j^{\prime}}\left(x\right) e^{-x}&= \delta^{j}_{j^{\prime}},
\end{align}
\end{subequations}
with $\delta_{n}^{m}$ the Kronecker delta. For simplicity, $T_{i0}$ is chosen as a constant in \eqref{eq:Hpj_expansions}, in contrast to \cite{Frei2020} where a dynamical bi-Maxwellian is used as the weight function for the Hermite and Laguerre polynomial basis. We note that the generalization of the projection in \eqref{eq:Hpj_expansions} onto a basis with anisotropic temperatures is straightforward. 

We define the DK velocity-space projection of a general phase-space function, $\chi$, on the $p$'th Hermite and $j$'th Laguerre polynomial, that is
\begin{subequations}
\begin{align}
    \projpj{pj}{\chi}_i&= \int_{-\infty}^{\infty} dv_{\|}\int_{0}^{\infty} d\mu \frac{B}{m_i}\int_0^{2\pi} d\theta F_{iDK}\frac{H_p L_j}{\sqrt{2^p p!}} \chi\left(\R, v_{\|}, \mu, \theta\right).
\intertext{And we further introduce the shorthand notation,}
    \projpj{}{\chi}_i&=\projpj{00}{\chi}_i.\label{eq:proj_DK}
\end{align}
\end{subequations}

We notice that in terms of the projector in \eqref{eq:proj_DK} and the gyromoments in \eqref{eq:DK_Npj_exp}, the DK ion density is given by
\begin{subequations}
\begin{align}
    N_{iDK}&= \projpj{}{1}_i=\N^{00}_{iDK},\label{eq:density_def}
\intertext{the DK ion parallel velocity by}
    U_{\| iDK}&=\projpj{}{v_{\|}}_i/N_{iDK},\label{eq:Upara_DK}
\intertext{the DK ion parallel pressure by}
    P_{\| iDK}&=m_i\projpj{}{\left(v_{\|}-U_{\| iDK}\right)^2}_i=T_{i0}\left(\sqrt{2}\N^{20}_{iDK}+N_{iDK}\right),
\intertext{and the DK ion perpendicular pressure by}
    P_{\perp i DK}&=\projpj{}{\mu B}_i=T_{i0}\left(N_{iDK}-\N^{01}_{iDK}\right).\label{eq:P_perp_def}
\intertext{Finally, from \eqref{eq:Upara_DK}, we have that}
    \N^{10}_{iDK}&=\sqrt{2}\projpj{}{s_{\| i}}_i=0.\label{eq:N10eq0}
\end{align}
\end{subequations}

Projecting \eqref{eq:Boltzmann_DKa} for $a=i$ on the Hermite Laguerre basis gives a gyromoment hierarchy
\begin{subequations}
\begin{align}
    &\frac{\partial}{\partial t} \N^{pj}_{iDK}  + \frac{\sqrt{2p}}{v_{Th i}} \N^{p-1j}_{iDK} \frac{\partial}{\partial t} U_{\| iDK}  + \frac{\sqrt{2p}}{v_{Th i}} \projpj{p-1 j}{\dot{\R}}_{i0}\cdot\nabla U_{\| iDK}      \nonumber\\
    &+ \nabla \cdot \projpj{pj}{ \dot \R}_{i0} - \frac{\sqrt{2p}}{v_{Th i}}  \projpj{p-1j}{ \dot v_\parallel}_{i0}=  \C^{pj}_{ii0}+\C_{ie 0}^{pj} + S^{pj}_{i0},\label{eq:moment_hierachy_i}
    \intertext{with}
    &\nabla \cdot \projpj{pj}{ \dot \R}_{i0} = \nabla_{\|}\projpj{pj}{v_{\|}}_i+\frac{1}{B}\poissonbracket{\phi_{DK}+\phi_{GK}}{\N^{pj}_{iDK}},\\
    &\projpj{p j}{\dot{\R}}_{i0}\cdot\nabla U_{\| iDK} = \projpj{pj}{v_{\|}}_i\nabla_{\|} U_{\| iDK}+\frac{\N^{pj}_{iDK}}{B}\poissonbracket{\phi_{DK}+\phi_{GK}}{U_{\| iDK}},\\
    &\projpj{pj}{ \dot v_\parallel}_i =-\frac{q_i}{m_i}\N^{pj}_{iDK} \nabla_{\|} \phi_{DK},
    \intertext{and}
    &\projpj{pj}{  v_\parallel}_i = \frac{v_{Th i}}{\sqrt{2}}\left(\sqrt{p+1}\N^{p+1 j}_{iDK}+\sqrt{p}\N^{p-1 j}_{iDK}\right)+U_{\| i DK}\N^{pj}_{iDK},
\end{align}
\end{subequations}
where we introduce the small-scale-averaged projection of the ion-ion collision operator, $\C^{pj}_{ii0}=\projpj{pj}{ C_{iiDK}/F_{iDK}}$, the ion-electron collision operator, $\C^{pj}_{ie0}=\projpj{pj}{ C_{ieDK} /F_{iDK}}$, and the ion source, $S^{pj}_{i0}=\projpj{pj}{ S_{iDK}/F_{iDK}}$.

The equation for $U_{\| i DK}$ is retrieved by setting $\left(p,j\right)=\left(1,0\right)$ in \eqref{eq:moment_hierachy_i}, that is
\begin{subequations}
\begin{align}
    m_i N_{iDK}\partial_t U_{\| i DK}&= -m_i N_{iDK}\left(U_{\| i DK}\nabla_{\|} U_{\| i DK}+\frac{1}{B}\poissonbracket{\phi_{DK}+\phi_{GK}}{U_{\| i DK}}\right)\nonumber\\
    &-\nabla_{\|} P_{\| iDK}-e \nabla_{\|}\phi_{DK}-F_{ei}+\frac{m_i}{\sqrt{2}} v_{Th i} S^{10}_{i0}+\mathcal{O}\left(m_i N_{iDK}c_s \omega \right),\label{eq:Upari}
\intertext{with}
    F_{ei}&=m_e\int d\bm v v_{\|} C_{ei}=-m_i\int d\bm v v_{\|} C_{ie},
\end{align}
\end{subequations}
due to conservation of momentum.

Explicitly inserting the relation for $\partial_t U_{\| i DK}$, \eqref{eq:Upari}, in \eqref{eq:moment_hierachy_i} we get
\begin{subequations}
\begin{align}
        \partial_t \N^{pj}_{iDK}&+ \nabla_{\|}\projpj{pj}{v_{\|}}_{i}+\sqrt{p}\projpj{p-1 j}{s_{\| i}}_{i} \nabla_{\|}U_{\| iDK}+\frac{1}{B}\poissonbracket{\phi_{DK}+\phi_{GK}}{\N^{pj}_{iDK}}\nonumber\\
        &- \frac{\sqrt{2p}\N^{p-1 j}_{i}}{N_{iDK}}\frac{1}{m_i v_{Thi}}\nabla_{\|} P_{\| i DK}=  \sum_{s}C_{is}^{pj} + S^{pj}_{i0}-\frac{\sqrt{p}\N^{p-1 j}_{iDK}}{N_{iDK}m_i v_{Th i}}S_U,\label{eq:moment_hierachy_DK}\\
        \projpj{pj}{s_{\|}}_i& =  \sqrt{p+1}\N^{p+1 j}_{iDK}+\sqrt{p}\N^{p-1 j}_{iDK}.
\end{align}
\end{subequations}

Finally, we use a long-wavelength Dougherty collision operator \cite{Dougherty1964} in \eqref{eq:moment_hierachy_DK} whose projection is given by \cite{frei2023fullf,mencke2025extended}
\begin{align}
    \mathcal{C}_{ii0}^{pj}=&\nu_i\left[-\left(p+2j\right)\N^{pj}_{iDK}+\left(T_{iDK}-1\right)\vphantom{\left(\sqrt{p\left(p-1\right)}\N^{p-2 j}_{iDK}-2j \N^{pj-1}_{iDK}\right)}\right.\nonumber\\
    &\left.\vphantom{-\left(p+2j\right)\N^{pj}_{iDK}+\left(T_i-1\right)} \times \left(\sqrt{p\left(p-1\right)}\N^{p-2 j}_{iDK}-2j \N^{pj-1}_{iDK}\right)\right],\label{eq:C_iDK}
\end{align}
with $\nu_{i}=\nu_0 1.38\sqrt{m_i/m_e}\left(T_{iDK}/T_{e0}\right)^{-3/2}$ and $\nu_0  = 4 \sqrt{2 \pi} e^4 N_{0}    \sqrt{m_e} \ln \lambda /[  3  m_i T_{e0}^{3/2} 1.96  \left(4\pi \epsilon_0\right)^2]$. 

\section{Gyrokinetic governing equations}\label{sec:GK_sys}
For the governing equations of the GK fields, we first develop the GK part of the electron distribution function in a similar way as the DK counterpart assuming an adiabatic response to the GK electrostatic potential. We then derive a moment hierarchy for the GK ion gyromoments, similarly as for the DK ion gyromoments. 

Focusing first on the electron species, we perturbatively solve the Boltzmann equation by expanding in orders of $\epsilon_m$. We first write down the order $\epsilon_m^{-1}\partial_t F_{eGK}$ of \eqref{eq:Boltzmann_GKa} with $a=e$, that is
\begin{align}
    v_{\|}\nabla_{\|} F_{eGK}-\frac{e}{m_e}\nabla_{\|}\phi_{DK}\partial_{v_{\|}} F_{eGK}-\frac{e}{m_e}\nabla_{\|}\phi_{GK}\partial_{v_{\|}} F_{eDK}=\C_{ee1}\left(F_{eDK},F_{eGK}\right)+\mathcal{L}_{ei}\left(F_{eGK}\right).\label{eq:Boltzmann_ad_egk}
\end{align}
Since \eqref{eq:Boltzmann_ad_egk} is linear in $F_{eGK}$, its general solution is given by
\begin{equation}
    F_{eGK}=\left(\frac{\delta N_{eGK}\left(\mathbf{R}_{\perp},t\right)}{N_{eDK}}+\frac{e\phi_{GK}}{T_{eDK}}\right)F_{eDK},\label{eq:FeGK}
\end{equation}
where $\delta N_{eGK}$ is a free parameter. We set $\delta N_{eGK}=0$, allowing us to close our system of equations. This corresponds to an adiabatic Boltzmann response for the GK part of the electron distribution function. 

For the ions, similarly to the DK system, we expand the GK part of the ion distribution function on a Hermite-Laguerre basis
\begin{subequations}
\begin{align}
    F_{iGK}\left(\mathbf{R},v_{\|}, \mu\right)&=\sum_{p,j=0}^{\infty}\N^{pj}_{iGK}\left(\R\right)\frac{H_{p}\left(s_{\| i}\right)L_{j}\left(x_i\right)}{\sqrt{2^{p}p!}} F_{Mi}\left(s_{\| i},x_i\right).
    \intertext{The GK velocity-space projection of a general phase-space function, $\chi$, is defined}
    \projpj{\delta pj}{\chi}_i&=\int_{-\infty}^{\infty} dv_{\|}\int_{0}^{\infty} d\mu \frac{B}{m_i}\int_0^{2\pi} d\theta F_{iGK}\frac{H_p L_j}{\sqrt{2^p p!}} \chi\left(\R, v_{\|}, \mu, \theta\right),
\intertext{where we additionally define,}
    \projpj{\delta }{\chi}_i&= \projpj{\delta 00}{\chi}_i.\label{eq:proj_GK}
\intertext{The first few moments of the GK distribution function can be expanded in terms of the GK and DK fluid quantities and gyromoments as follows}
\projpj{\delta}{1}_i&= \N^{00}_{iGK}=N_{iGK},\\
    \projpj{\delta}{v_{\| }}_i&=U_{\| iDK} N_{iGK} +\frac{v_{Th i}}{\sqrt{2}}\N^{10}_{iGK}=N_{iDK} U_{\| iGK}+U_{\| iDK} N_{iGK},\label{eq:N_aUpara_GK}\\
    m_i\projpj{\delta}{\left(v_{\|}-U_{\| i DK}\right)^2}_i&-2m_i\projpj{}{\left(v_{\|}-U_{\| iDK}\right)U_{\| iGK}}_i \nonumber\\
    &=T_{i0}\left(\sqrt{2}\N^{20}_{iGK}+ N_{iGK}\right)=P_{\| i GK},\\
    \projpj{\delta}{\mu B}_i&=T_{i0}\left( N_{iGK}- \N^{01}_{iGK}\right)=P_{\perp iGK},
\intertext{and, from \eqref{eq:N_aUpara_GK}, it follows that}
     U_{\| iGK}&=\frac{v_{Th i}}{\sqrt{2}}\frac{\N^{10}_{iGK}}{N_{iDK}}.
\end{align}
\end{subequations}
In order to derive the evolution equations for $\N^{pj}_{iGK}$, we project \eqref{eq:Boltzmann_GKa} with $a=i$ on the Hermite-Laguerre basis, thus obtaining
\begin{subequations}
    \begin{align}
    &\frac{\partial}{\partial t} \N^{pj}_{iGK}  + \frac{\sqrt{2p}}{v_{Th i}} \projpj{p-1 j}{\dot{\R}}_{i1}\cdot\nabla U_{\| iDK}     + \nabla \cdot \projpj{pj}{ \dot \R}_{i1} - \frac{\sqrt{2p}}{v_{Th i}}  \projpj{p-1j}{ \dot v_\parallel}_{i1} \nonumber\\
    & + \frac{\sqrt{2p}}{v_{Th i}} \N^{p-1j}_{iGK} \frac{\partial}{\partial t} U_{\| iDK}  + \frac{\sqrt{2p}}{v_{Th i}} \projpj{\delta p-1 j}{\dot{\R}}_{i1}\cdot\nabla U_{\| iDK}     + \nabla \cdot \projpj{\delta pj}{ \dot \R}_{i1} - \frac{\sqrt{2p}}{v_{Th i}}  \projpj{\delta p-1j}{ \dot v_\parallel}_{i1} \nonumber\\
    &=  \C^{pj}_{ii1}+\C_{ie1}^{pj} + \delta S^{pj}_{i},\label{eq:moment_hierachy_gk_i}
\intertext{where}
    &\projpj{p j}{\dot{\R}}_{i1}\cdot\nabla U_{\| }=\frac{1}{B}\projpj{pj}{\bm b\times \nabla \left(\gyroavg{\psi}-\gyroavg{\phi_{GK}}\right)}_i\cdot\nabla U_{\| }+\frac{1}{B}\projpj{pj}{\bm b\times\nabla\gyroavg{\widetilde{\phi_{GK}}}}_{i}\cdot\nabla U_{\| },\\
    &\nabla \cdot \projpj{pj}{ \dot \R}_{i1}=\frac{1}{B}\nabla\cdot\projpj{pj}{\bm b\times \nabla \left(\gyroavg{\psi}-\gyroavg{\phi_{GK}}\right)}_i+\frac{1}{B}\nabla\cdot\projpj{pj}{\bm b\times\nabla\gyroavg{\widetilde{\phi_{GK}}}}_{i},\\
    &\projpj{pj}{ \dot v_\parallel}_{i1}= -\frac{q_i}{m_i}\projpj{pj}{\nabla_{\|} \gyroavg{\phi_{GK}}}_i,\\
    &\N^{p-1j}_{iGK} \frac{\partial}{\partial t} U_{\| }  +  \projpj{\delta p-1 j}{\dot{\R}}_{i1}\cdot\nabla U_{\| }-\projpj{\delta p-1j}{ \dot v_\parallel}_{i1}=\frac{v_{Th i}}{\sqrt{2}}\projpj{\delta p-1 j}{s_{\| i}}_i\nabla_{\|}U_{\|}\nonumber\\
    &-\frac{ \N^{p-1 j}_{iGK}}{N_{iDK}}\left(\frac{1}{m_i}\nabla_{\|} P_{\| i}-\frac{S_U}{m_i}\right),
\intertext{and}
    &\nabla \cdot \projpj{\delta pj}{ \dot \R}_{i1}=\nabla_{\|}\projpj{\delta pj}{v_{\|}}_i+\frac{1}{B}\poissonbracket{\phi_{DK}}{ \N^{pj}_{iGK}}+\frac{1}{B}\nabla\cdot\projpj{\delta pj}{\bm b\times \nabla \gyroavg{\phi_{GK}}}_i.
\end{align}
\end{subequations}
In \eqref{eq:moment_hierachy_gk_i} we introduce the linearized (around $F_{iDK}$ and $F_{eDK}$) ion-ion and ion-electron collision operator, $\C^{pj}_{ii1}$ and $\C^{pj}_{ie1}$, and the small-scale source term, $\delta S^{pj}_1=\int d\mathbf{v} H^{pj}\left(S_i-\left\langle S_i\right\rangle_{DK}\right)$. The expressions of the projection of the gyroaveraged quantities involving $\gyroavg{\psi}$ and $\gyroavg{\phi_{GK}}$ are reported in \appref{app:proj_gyro_avg}. We use the GK Dougherty collision operator in \eqref{eq:moment_hierachy_gk_i} \cite{Frei2020}
\begin{align}
    \mathcal{C}_{ii1}^{pj}=&\nu_{i0}\left[-\left(p+2j\right)\N^{pj}_{iGK}+\tau_i T_{iDK}\nabla_{\perp}^2\N^{pj}_{iGK} +\left(T_{iDK}-1\right)\vphantom{\left(\sqrt{p\left(p-1\right)}\N^{p-2 j}_{iGK}+T_{i}\nabla_{\perp}^2-2j \N^{pj-1}_{iGK}\right)}\right.\nonumber\\
    &\left.\vphantom{-\left(p+2j\right)\N^{pj}_{iGK}+\left(T_{iDK}-1\right)} \times \left(\sqrt{p\left(p-1\right)}\N^{p-2 j}_{iGK}-2j \N^{pj-1}_{iGK}\right)\right],\label{eq:C_iGK}
\end{align}
where we approximate $T_i = T_{iDK}$.

\section{Field equations}\label{sec:field_equations}
In order to close our system of equations, we now determine the governing equations for $\phi_{DK}$ and $\phi_{GK}$. In the process, we will also determine a relation between $N_{iDK}$ and $N_{eDK}$ which allows us to avoid solving both \eqref{eq:dt_Ne} and \eqref{eq:moment_hierachy_DK} with $\left(p,j\right)=\left(0,0\right)$. For this purpose, we perform the variation of the action,
\begin{equation}
    A=\sum_{a} \int dt \int d\mathbf{x}\int d \mathbf{v}F_a\Gamma_a,\label{eq:Action_poisson}
\end{equation}
with respect to $\phi_{GK}$ where quasineutrality is imposed by neglecting the field functional action term \cite{Frei2020}, and $\Gamma_i$ and $\Gamma_e$ are defined in Eqs (\ref{eq:Gammai}) and (\ref{eq:gamma_e}), respectively. We note that, the variation with respect to $\phi_{DK}$, yields an identical result in the long wavelength limit, being accurate only to order $\left(\rho_s k_{\perp GK}\right)^2$ since we only allow $\phi_{DK}$ to vary on long scales.

The variation of the action, \eqref{eq:Action_poisson}, with respect to $\phi_{GK}$ gives the following Poisson equation,

\begin{subequations}
\begin{align}
    \frac{\Bpar}{B}N_{iDK}&-N_{eDK}+\frac{1}{2m_i\Omega_i^2}\nabla_{\perp}^2 P_{\perp i DK}+\varrho_{i}^*- N_{eGK}\nonumber\\
    &+\nabla_{GK}\left(\gyroavg{\phi_{GK}}\right)=\mathcal{O}\left(n_e\epsilon_{\perp}^3,n_e\epsilon_{\delta}\epsilon_{\perp},n_e\epsilon_{\delta}^2\right),\label{eq:Poisson_full}
\intertext{where we introduce the GK Poisson operator}
    \nabla_{GK}\left(\gyroavg{\phi_{GK}}\right)&=2\pi \int_{-\infty}^{\infty} dv_{\|}\int_{0}^{\infty} d\mu \frac{B}{m_i} \left( F_{iDK} \frac{e}{B}\partial_\mu \gyroavgadj{\gyroavg{\phi_{GK}}}\right),
\intertext{and the ion GK particle density}
    \varrho_{i}^*&=2\pi \int_{-\infty}^{\infty} dv_{\|}\int_{0}^{\infty} d\mu \frac{B}{m_i} \gyroavgadj{F_{iGK}},\label{eq:varrho}\\
\end{align}
\end{subequations}
and where the adjoint gyroaverage operator has been defined as
\begin{subequations}
\begin{align}
    \gyroavgadj{f}&=\frac{1}{2\pi}\int_{0}^{2\pi}d\theta \int d\mathbf{R}\delta\left(\mathbf{R}+\boldsymbol{\rho}-\mathbf{x}\right) f\left(\mathbf{R},\mu,v_{\|}, t\right)\\
    &=\sum_{n=0}^{\infty}\left(\frac{2\mu}{q_a\Omega_a}\right)^{n}\frac{\nabla_{\perp}^{2n}}{2^n \left(n!\right)^2}f\left(\boldsymbol{x}, {\mu}, {v}_{\|}, t\right),
\end{align}
\end{subequations}
valid for  constant $B$ (in this case the gyroaverage operator is self-adjoint).

We impose that \eqref{eq:Poisson_full} is satisfied independently order by order, up to leading order in $\epsilon_{\perp}$ and $\epsilon_{\delta}$. For this purpose, we introduce the higher order DK densities such that
\begin{align}
    N_{aDK}=N_{aDK0}+N_{aDK1},
\end{align}
with $N_{aDK1}\sim \epsilon_{\perp}^2 N_{aDK}$. This expansion removes the need of evaluating the Pfirsch-Kaufman polarization \cite{pfirsch1985local,kaufman1986electric,Frei2020}. At order $N_e$, \eqref{eq:Poisson_full} reads
\begin{equation}
    N_{iDK 0}-N_{eDK 0} =0,\label{eq:Poisson00}
\end{equation}
thus we impose $N_{iDK 0}=N_{eDK 0}=N_{DK}$. By taking the time derivative of \eqref{eq:Poisson00}, using \eqref{eq:dt_Ne} and \eqref{eq:moment_hierachy_i} with $\left(p,j\right)=\left(0,0\right)$, we relate the electron and ion parallel velocities as follows
\begin{equation}
    \nabla_{\|}\int d\bm v v_{\|}\left(F_{iDK}-F_{eDK}\right)\equiv \nabla_{\|}J_{\|}=\mathcal{O}\left(n_e\Omega_i\epsilon\epsilon_{\perp}^3\right).\label{eq:cond_Jpar}
\end{equation}

In order to satisfy \eqref{eq:cond_Jpar}, we assume $J_{\|}=\mathcal{O}\left(c_{s}n_e\epsilon_{\perp}^2\right)$. This means that to the lowest order in $\epsilon_{\perp}$, the electrons and ions move with one fluid velocity, $U_{\|}$. Therefore, we define 
\begin{equation}
    m_i N_{DK}U_{\|}\equiv 2\pi \int_{-\infty}^{\infty} dv_{\|}\int_{0}^{\infty} d\mu \frac{B}{m_i} v_{\|}\left(m_e F_{eDK}+ m_i F_{iDK}\right)+\mathcal{O}\left(c_{s}\epsilon_{\perp}^2\right).\label{eq:Upar_def}
\end{equation}
We use that to the lowest order in $\epsilon_{m}$, $U_{\| i DK}=U_{\| e DK}+\mathcal{O}\left(c_{s}\epsilon_\perp^2,c_{s}\epsilon_{m}^2\right)$ and we determine the governing equation for $U_{\|}$ as a total momentum equation, adding $m_e N_{DK}\partial_t U_{\| e DK}+m_i N_{DK}\partial_tU_{\| i DK}$, as it is done in the drift-reduced Braginskii model \cite{braginskii1965transport,Zeiler1997,FelixParraBrag}. The moment with $m_e v_{\|}$ of \eqref{eq:Boltzmanne1} gives
\begin{subequations}
\begin{align}
    m_e N_{DK}\partial_t U_{\| e DK}&= eN_{DK}\nabla_{\|}\phi_{DK}-\nabla_{\|}P_{eDK}+F_{ei}+\frac{m_e}{\sqrt{2}} v_{Th e} S^{10}_{e0}\nonumber\\
    &+\mathcal{O}\left(m_i N_{DK}c_s \Omega_i\epsilon \epsilon_\perp^3,m_eN_{DK}c_s \Omega_i\epsilon \epsilon_\perp\right),\label{eq:dt_upare_low}
\intertext{and combining Eqs. (\ref{eq:Upari}) and (\ref{eq:dt_upare_low}) gives}
    m_iN_{DK}&\left(\partial_{t} U_{\| }+ U_{\| }\nabla_{\|} U_{\| }+\frac{1}{B}\poissonbracket{\phi_{DK}+\phi_{GK}}{U_{\| }}\right)\nonumber\\
     +\nabla_{\|} P_{\| iDK}&+\nabla_{\|} P_{\| eDK} =S_U+\mathcal{O}\left(m_i N_{DK}c_s \Omega_i\epsilon \epsilon_\perp^3,m_eN_{DK}c_s \Omega_i\epsilon \epsilon_\perp\right).\label{eq:Upar}
\end{align}
\end{subequations}
Thus $\partial_t U_{\|}$ is independent of the collision frequencies and we avoid to numerically satisfy momentum conservation, $m_e\int d\bm v v_{\|}C_{ei}^{10}=-m_i\int d\bm v v_{\|} C_{ie}^{10}$, at high collisionality. In \eqref{eq:Upar}, we define 
\begin{equation}
    S_U=\frac{1}{\sqrt{2}}\left(m_i v_{Th i} S^{10}_{i0}+m_e v_{Th e}S^{10}_{e0}\right).
\end{equation}
We evolve $U_{\|}$ through \eqref{eq:Upar} and impose $U_{\|i}=U_{\|e}=U_{\|}$ in Eqs. (\ref{eq:dt_Ne}), (\ref{eq:dt_Te}), and (\ref{eq:moment_hierachy_DK}) in our simulations. Furthermore, \eqref{eq:dt_Ne} is identical to \eqref{eq:moment_hierachy_DK} with $\left(p,j\right)=\left(0,0\right)$ so we only evolve \eqref{eq:dt_Ne} and impose $\N^{00}_{iDK}=N_{eDK}=N_{DK}$. Finally, as an aside, we note that the implicit condition $J_{\|}=\mathcal{O}\left(n_ec_{s}\epsilon_{\perp}^2,n_ec_{s}\epsilon_{m}^2\right)$ implies parallel force balance given as
\begin{align}
    &\frac{e}{m_e}\nabla_{\|}\phi_{DK}-\frac{1}{m_e}\nabla_{\|}P_{eDK}+\frac{v_{The}}{\sqrt{2}}S_{e0}^{10}-\frac{1}{m_e}F_{ei}-\frac{v_{Thi}}{\sqrt{2}}S_{i0}^{10}+\frac{1}{m_i}\nabla_{\|}P_{i\| DK}\nonumber\\
    &=\mathcal{O}\left( N_{DK}c_s\Omega_i\epsilon\epsilon_{\perp}^3,N_{DK}c_{s}\Omega_i\epsilon\epsilon_{\perp}\epsilon_{m}^2\right).
\end{align}
In other words, in order for quasi-neutrality to be satisfied to a certain order in $\epsilon_\perp$, the acceleration of ions and electrons must balance to the same order.

To obtain an equation for $\phi_{DK}$, we have to consider the order $\epsilon_\perp^2$ terms of \eqref{eq:Poisson_full}, that is
\begin{equation}
    N_{eDK1}-N_{iDK1}=\frac{1}{B \Omega_i }N_{DK}\nabla_{\perp}^2\phi_{DK}+\frac{1}{2m_i\Omega_i^2}\nabla_{\perp}^2 P_{\perp i DK}\label{eq:Poisson20}.
\end{equation}
In order to avoid a closure problem in $\epsilon_{\perp}$, which arises by evolving $N_{eDK1}$ and $N_{iDK1}$ independently, we define the vorticity,
\begin{align}
    \Omega &\equiv N_{eDK1}-N_{iDK1}+\frac{1}{B\Omega_i}\nabla_{\perp} N_{DK}\cdot\nabla_{\perp} \phi_{DK}+\frac{1}{2m_i\Omega_i^2}\nabla_{\perp}^2 P_{\perp i DK}\label{eq:Omega_def}
\intertext{as is custom in fluid approaches \cite{Zeiler1997}. Using \eqref{eq:Poisson20}, $\Omega$ can also be written as}
    \Omega&=\frac{1}{B\Omega_i}\nabla\cdot\left(N_{DK}\nabla_{\perp}\phi_{DK}+\frac{1}{q_i \Omega_i}\nabla_{\perp}P_{\perp i DK}\right).
\end{align}
The addition of the pressure contribution in \eqref{eq:Omega_def} makes $\Omega$ independent of collisions. Indeed, the particle density is independent of collision such that $\partial_t n_i$ with \[ 
 n_i\equiv 2\pi \int_{-\infty}^{\infty} dv_{\|}\int_{0}^{\infty} d\mu \frac{B}{m_i}\gyroavgadj{F_{iDK}}=N_{DK}+N_{iDK 1}+\frac{1}{2m_i\Omega_i^2}\nabla_{\perp}^2 P_{\perp i DK}+\mathcal{O}\left(N_{DK} \epsilon_{\perp}^4\right)
\]
is independent of collisions, but not $\partial_t N_{iDK 1}$ on its own.




We differentiate each term of \eqref{eq:Omega_def} with respect to $t$. For this purpose, we note that, by using the Boltzmann equation, \eqref{eq:Boltzmann_a}, averaged over small-scale fluctuations but including $\epsilon\epsilon_{\perp}^3$ terms, we have
\begin{subequations}
    \begin{align}
    \partial_t N_{iDK1}&=-\nabla_{\|}\left(U_{\|}N_{iDK 1}+N_{DK}U_{i\| DK 1}\right)-\frac{1}{B}\poissonbracket{\phi_{DK}+\phi_{GK}}{\N_{iDK 1}}&\nonumber\\
    &-\frac{1}{B}\poissonbracket{\phi_{DK1}+\phi_{GK1}}{\N_{DK}}+S_{N1}+\frac{1}{B\Omega_i}S_{N}\nabla_\perp^2\phi_{DK}-\frac{1}{B\Omega_i}\partial_t\left(N_{DK}\nabla_\perp^2\phi_{DK}\right)\nonumber\\
    &+\frac{1}{B^2\Omega_i}\nabla_\perp\cdot\left(N_{DK}\nabla_\perp \partial_t\phi_{DK}\right)-\frac{1}{B\Omega_i}\nabla_{\perp}^2\phi_{DK} \nabla_{\|}\left(N_{DK}U_{\|}\right)\nonumber\\
    &-\frac{m_i}{2B^3}\poissonbracket{\left|\nabla_\perp \phi_{DK}\right|^2}{N_{DK}}-\frac{1}{2m_i\Omega_i^2B}\poissonbracket{\nabla_\perp^2\phi_{DK}}{P_{\perp i DK}}\nonumber\\
    &+\frac{1}{B\Omega_i}\nabla_{\|}\nabla_\perp \phi_{DK}\cdot\nabla_{\perp}\left(\N_{DK}U_{\|}\right)+\frac{1}{2m_i\Omega_i^2}\nabla_{\perp}^2\projpj{}{\mu B \left(S_i+C_{ii}\right)/F_{iDK}},\label{eq:Nidk1}\\
\intertext{where the 1 subscript indicates order $\epsilon_\perp^2$ corrections. The last term in \eqref{eq:Nidk1} comes from the sources and collisions acting on the particle position and not gyrocenter position. For the time derivative of the higher order electron density, we get}
   \partial_t N_{eDK1}&=-\nabla_{\|}\left(U_{\|}N_{eDK 1}+N_{DK}U_{e\| DK 1}\right)-\frac{1}{B}\poissonbracket{\phi_{DK}+\phi_{GK}}{\N_{iDK e}}&\nonumber\\
    &-\frac{1}{B}\poissonbracket{\phi_{DK1}+\phi_{GK1}}{\N_{DK}}+S_{N1},\label{eq:Nedk1}
\intertext{using \eqref{eq:moment_hierachy_DK} and \eqref{eq:P_perp_def}, the time derivative of the pressure term in \eqref{eq:Omega_def} is given by}
    \partial_{t}\nabla_\perp^2& P_{\perp i DK}=-\nabla_{\|}\nabla_{\perp}^2\left(P_{\perp i DK}U_{\|}+Q_{\perp i DK}\right)\nonumber\\
    &-\frac{1}{B}\nabla_{\perp}^2\poissonbracket{\phi_{DK}+\phi_{GK}}{P_{\perp i DK}}+\nabla_{\perp}^2\projpj{}{\mu B \left(S_i+C_{ii}\right)/F_{iDK}},\label{eq:nablaPiDK}\\
\intertext{with $Q_{\perp i DK}=\projpj{}{v_{\|}\mu B}_i=-v_{Th i}T_{i0}\N^{11}_{iDK}/\sqrt{2}$. Adding Eqs. (\ref{eq:Nidk1}), (\ref{eq:Nedk1}), and (\ref{eq:nablaPiDK}) while using \eqref{eq:Omega_def}, we notice that all terms including a time derivative combine to $\nabla\cdot\left(\partial_tN_{DK}\nabla_\perp \phi_{DK}\right)$. Finally, using \eqref{eq:dt_Ne}, we get}
    \partial_t\Omega&=\nabla_{\|}\left(J_{\|}-\left(N_{eDK1}-N_{iDK1}\right)U_{\|}-\frac{1}{2m_i\Omega_i^2}\nabla_{\perp}^2\left(P_{\perp i DK}U_{\|}+Q_{\perp i DK}\right)\right)\nonumber\\
    &-\frac{1}{B}\poissonbracket{\phi_{DK}+\phi_{GK}}{N_{eDK1}-N_{iDK1}}-\frac{1}{2m_i\Omega_i^2B}\nabla_{\perp}^2\poissonbracket{\phi_{DK}+\phi_{GK}}{P_{\perp i DK}}\nonumber\\
    &-\frac{1}{B\Omega_i}\nabla_\perp\cdot\left[\nabla_\perp \phi_{DK} \left(\nabla_{\|}\left(U_{\|}N_{DK}\right)+\frac{1}{B}\poissonbracket{\phi_{DK}+\phi_{GK}}{N_{DK}}-S_N\right)\right]\nonumber\\
    &+\frac{1}{B\Omega_i}\nabla_\perp \partial_t \phi_{DK}\cdot\nabla_\perp N_{DK}+\frac{m_i}{2B^3}\poissonbracket{\left|\nabla_\perp \phi_{DK}\right|^2}{N_{DK}}\nonumber\\
    &-\frac{1}{2m_i\Omega_i^2B}\poissonbracket{\nabla_\perp^2\phi_{DK}}{P_{\perp i DK}}-\frac{1}{B\Omega_i}\nabla_\perp\cdot\left(N_{DK}\nabla_\perp \partial_t\phi_{DK}\right)\nonumber\\
    &+\frac{1}{B\Omega_i}\left(N_{DK}\nabla_\perp^2\partial_t\phi_{DK}\right)+\frac{1}{B\Omega_i}S_N\nabla_\perp^2 \phi_{DK},
\intertext{with $J_{\|}=N_{DK}\left(U_{\| i DK 1}-U_{\| e DK 1}\right)$. Finally, through tedious elementary vector calculus and algebra, we get}
    \partial_t\Omega&=\nabla_{\|} J_{\|}-\nabla_{\|}\left(\Omega U_{\| iDK}\right)-\frac{1}{B}\poissonbracket{\phi_{DK}+\phi_{GK}}{\Omega}-\frac{1}{2m_i\Omega_i^2}\nabla_{\|}\left(P_{\perp i DK} \nabla_{\perp}^2 U_{\| i DK}\right)\nonumber\\
    &-\frac{1}{B}\poissonbracket{\nabla_{\perp}\left(\phi_{DK}+\phi_{GK}\right)\cdot}{\boldsymbol{\omega} }-\nabla_{\|}\left(\boldsymbol{\omega}\cdot\nabla_{\perp}U_{\| i DK}\right)+\frac{1}{B\Omega_i}\nabla_{\perp}S_N\cdot\nabla_{\perp} \phi_{DK}\nonumber\\
    &+\frac{1}{B^2\Omega_i}\poissonbracket{\nabla_\perp \phi_{DK}\cdot}{N_{iDK}\nabla_\perp \phi_{GK}}+\frac{1}{2m_i\Omega_i^2}\poissonbracket{\nabla_\perp^2\phi_{GK}}{P_{\perp i DK}}\nonumber\\
    &-\frac{1}{2m_i\Omega_i^2}\nabla_{\|}\left(\nabla_{\perp}^2 Q_{\perp i DK}\right),\label{eq:dt_Omega}\\
    \intertext{with}
    \boldsymbol{\omega}&=\frac{1}{B\Omega_i}N_{DK}\nabla_{\perp}\phi_{DK}+\frac{1}{m_i\Omega_i^2}\nabla_{\perp}P_{\perp i DK},
    \intertext{and as a consequence}
    \Omega&=\nabla_{\perp}\cdot\boldsymbol{\omega}.\label{eq:Omega_impl}
\end{align}
\end{subequations}
Since \eqref{eq:dt_Omega} requires $J_{\|}$ to first non-vanishing order, we derive an equation for $J_{\|}$, describing it as a dynamical field. Noticing that
\begin{subequations}
\begin{align}
    J_{\|}&\equiv N_{DK}\left(U_{\| i DK}-U_{\| e DK}\right)\nonumber\\
    &=N_{DK}\left(U_{\| i DK1}-U_{\| e DK1}\right)\sim \mathcal{O}\left( N_{DK}c_s\epsilon_{\perp}^2,N_{DK}c_s\epsilon_{m}^2 \right),\label{eq:Jpar_deff}
\intertext{we remark that that $J_{\|}$ is the DK current density and does not include the GK distribution functions. Taking the derivative of \eqref{eq:Jpar_deff}, we get}
    \partial_t J_{\|}&+\frac{1}{B}\poissonbracket{\phi_{DK}+\phi_{GK}}{J_{\|}}+\nabla_{\|}\left(J_{\|} U_{\|}\right)+J_{\|}\nabla_{\|} U_{\|}+\frac{e}{m_e}N_{DK}\nabla_{\|} \phi_{DK}+\frac{1}{m_i}\nabla_{\|} P_{\| iDK}\nonumber\\
    &+\frac{1}{m_e}\nabla_{\|} \left(N_{DK}T_{eDK}\right)=\frac{S_N}{N_{DK}}J_{\|}-\nu_{\|}\frac{N_{DK}m_i}{N_0 m_e} J_{\|}-0.71\frac{N_{DK}}{m_e}\nabla_{\|}T_{eDK}+S_J,\label{eq:dt_J}
\end{align}
\end{subequations}
having introduced the source term, $S_j\equiv \int d\mathbf{v} \left(\left(v_{\|}-U_{\| i DK}\right)S_i-\left(v_{\|}-U_{\| e DK}\right)S_{e}\right)$ and the Spitzer resistivity $\nu_{\|}=\nu_0\left(T_e/T_{e0}\right)^{-3/2}$.

Eqs. (\ref{eq:dt_Omega}) and (\ref{eq:dt_J}) are equivalent to the vorticity and current density equation for the drift-reduced Braginskii model \cite{Zeiler1997} except for the addition of the source term, $\sim \nabla_{\perp} S_N$, arising from considering the pressence of sources, and the terms $\sim \nabla_{\|}\left(P_{\perp i DK}\nabla_{\perp}^2U_{\| i DK}+\nabla_\perp^2Q_{\perp i DK}\right)$ in \eqref{eq:dt_Omega} due to the distinction between particle and gyrocenter parallel velocity. In this work, we use the gyrocenter parallel velocity while in drift-reduced Braginskii equations, the particle parallel velocity is used. We remark that the term $ T_{eDK}\nabla_{\|} J_{\|}/N_{eDK}\sim \epsilon_\perp^2 \partial_t T_e$ in \eqref{eq:dt_Te} is of higher order, therefore we choose to neglect it in our simulations.

Finally, in order to obtain an equation for $\phi_{GK}$, we evaluate the order $\epsilon_{\delta}$ terms in \eqref{eq:Poisson_full} and insert $N_{eGK}$ from \eqref{eq:FeGK}
\begin{equation}
    \varrho_{i}^*=\frac{e\phi_{GK}}{T_{eDK}}N_{DK}-\frac{q_i}{B}\projpj{}{\partial_\mu \gyroavgadj{\gyroavg{\phi_{GK}}}}_i.\label{eq:Poisson_GK}
\end{equation}
We note that the left hand side of \eqref{eq:Poisson_GK} only depends on $\N^{pj}_{iGK}$ while the right hand side is an operator, depending on the DK fields, acting on $\phi_{GK}$. In the limit where the DK distribution functions are constant Maxwellians, \eqref{eq:Poisson_GK} reduces to the standard $\delta$-f GK Poisson equation
\begin{equation}
    \varrho_{i}^*=eN_{DK}\left[\frac{1}{T_{eDK}}+\frac{1}{T_{iDK}}-\frac{1}{T_{iDK}}\exp\left(-\frac{T_{iDK}k_{\perp}^2}{m_i\Omega_i^2}\right)I_{0}\left(\frac{T_{iDK}k_{\perp}^2}{m_i\Omega_i^2}\right)\right]\phi_{GK},
\end{equation}
with $I_0$ the modified Bessel function of order zero.
\section{Simulation setup and numerical implementation}\label{sec:Num_impl}
In order to implement the system of Eqs. (\ref{eq:dt_Ne}), (\ref{eq:dt_Te}), (\ref{eq:moment_hierachy_i}) for $\left(p,j\right)\neq \left(0,0\right)$ and $\left(p,j\right)\neq\left(1,0\right)$, (\ref{eq:moment_hierachy_gk_i}), (\ref{eq:Upar}), (\ref{eq:dt_Omega}), (\ref{eq:Omega_impl}), (\ref{eq:dt_J}),  and (\ref{eq:Poisson_GK}), closed by enforcing $\N^{00}_{iDK}=N_{eDK}=N_{DK}$, $\N^{10}_{iDK}=0$, and $U_{\| i DK}=U_{\| e DK}=U_{\|}$, we normalize time, $t$, to $ R / c_{s0}$ ($c_{s0} = \sqrt{T_{e0} / m_i}$ is the ion sound speed at the constant reference electron temperature, $T_{e0}$, and $R$ is the characteristic length that indicates the size of the plasma chamber in the direction perpendicular to $\bm B$). The electrostatic potentials, $\phi_{DK}$ and $\phi_{GK}$, are normalized to $ T_{e0} / e$, the parallel (to $\bm B$) spatial scale to $R$, and the perpendicular ones to $\rho_{s0} = c_{s0} / \Omega_i$. The densities, $N_{DK}$ and $N_{GK}$, are normalized to the constant reference density $N_{0}$, the parallel fluid velocities, $U_{\parallel }$, to $c_{s0}$, the electron and ion temperatures, $T_e$, $T_{\| i}$, and $T_{\perp i}$, to $T_{e0}$, $T_{i0}$, and $T_{i0}$, respectively. Finally, we normalize the parallel current density, $J_{\|}$, to $N_{0}c_{s0}$.

The domain we consider is a cuboid with sides $L_x$, $L_y$, and $L_z$. We introduce the spatial coordinates $(x,y,z)$, where $z$ is parallel to the magnetic field of the device, and $x$ and $y$ are perpendicular to $z$, forming a right-handed coordinate system.

To evolve the DK quantities, $\N^{pj}_{iDK}$, $U_{\|}$, $N_{DK}$, $T_{eDK}$, $\phi_{DK}$, $\Omega$ and $J_{\|}$, we use a finite-difference scheme similar to the one described in Ref. \cite{frei2023fullf,mencke2025extended}. The $x$ direction is discretized in $N_{x}$ equally sized grid points such that the grid spacing is $\Delta x=L_x/N_x$. An equivalent discretization is used for $y$ and $z$. Similarly to the GBS code \cite{Giacomin2022}, derivatives are calculated with a fourth order central difference scheme, except for the Poisson bracket operator, $\left[ f,g\right]$, which is evaluated using the fourth-order Arakawa algorithm \cite{Arakawa1997}. A staggered grid approach \cite{Paruta2018} is used in the $z$ direction with $U_{\|}$, $J_{\|}$, and the $\N^{pj}_{iDK}$ with odd $p$ being evolved on grid points shifted by a distance $\Delta z/2$ with respect to the grid points for $N_{DK}$, $T_{eDK}$, $\phi_{DK}$, $\Omega$, and $\N^{pj}_{iDK}$ with even $p$.

To evolve the GK fields, $\N^{pj}_{iGK}$ and $\phi_{GK}$, we use a spectral method in the $(x,y)$ plane, and finite differences along $z$. More precisely, we evolve $N_{kx}\times \lfloor N_{ky}/2+1 \rfloor$ Fourier modes with wavevectors $k_x$ and $k_y$ with spacing $\Delta k_x= 2\pi/L_x$ and $\Delta k_y= 2\pi/L_y$, respectively. Non-linear terms are evaluated in real space by inverse Fourier transforming the fields, multiplying them, and Fourier transforming back. All Fourier transforms are performed with the FFTW library \cite{FFTW05}. We note that $\phi_{GK}$ and $\N^{pj}_{iGK}$ with even $p$ are evolved on the same $z$-coordinates as $\phi_{DK}$, while the $\N^{pj}_{iGK}$ with odd $p$ are evolved on the same $z$-coordinates as $U_{\|}$. Finally, all fields are antialiased using the $2/3$ rule \cite{Hoffmann2023,Hoffmann2023b,orszag1971elimination}. 

For the DK and GK gyromoments, we close the hierarchy using a simple truncation approach. Thus, we evolve $\N^{pj}_{iDK}$ with $p$ from $0$ to $P_{DK}$ and with $j$ from $0$ to $J_{DK}$ for the DK moments. Similarly, for the GK moments, we evolve $\N^{pj}_{iGK}$ with $p$ from $0$ to $P_{GK}$ and with $j$ from $0$ to $J_{GK}$.

Similarly to \cite{mencke2025extended}, a fifth-order adaptive time-step Runge-Kutta scheme is used to advance all fields \cite{press19922numerical}. Equation (\ref{eq:Omega_impl}) is solved for $\phi_{DK}$ using an approach similar to the one used in the two-fluid GBS code \cite{Giacomin2022}. The discrete operator of the $\nabla\cdot\left(N_{DK}\nabla \phi_{DK} \right)$ term is constructed at each time step and the linear system is solved for $\phi_{DK}$ using the PETSCs library \cite{petsc-web-page}. Equation (\ref{eq:Poisson_GK}) is solved for $\phi_{GK}$ using an iterative approach. Since the electron adiabatic response generally dominates the right hand side of \eqref{eq:Poisson_GK}, for each $z$ plane we solve
\begin{align}
    \phi_{GK}^{n+1}&=\frac{T_{eDK}}{N_{eDK}}\left(\rho_i^*+\frac{q_i}{B}\projpj{}{\partial_\mu \gyroavgadj{\gyroavg{\phi_{GK}^{n}}}}_i\right)\nonumber\\
    \text{while: }& \max_{k_x,k_y}\left|\phi_{GK}^{n+1}-\phi_{GK}^{n}\right|>\eta_{GK},
\end{align}
and initialize $\phi_{GK}^{0}$ using $\phi_{GK}$ from the previous timestep. The parameter $\eta_{GK}>0$ is the error threshold chosen small enough not to impact the results (typically $\eta_{GK}\sim 10^{-7}$).


For stability reasons, numerical diffusion is added to the right-hand side of all equations for all evolved DK fields, $f$, that is
\begin{align}
D(f) = \eta_\perp \left(\partial_{x}^2 +  \partial_{y}^2\right) f + \eta_z \partial_{z}^2 f.
\end{align}

The numerical parameter $\eta_{\perp}>0$ and $\eta_{z}>0$ are chosen as small as possible to guarantee the numerical stability of the simulations, without significantly affecting the results. For the GK field, we do not add numerical diffusion in the perpendicular plane, i.e. $\eta_\perp=0$. However, we do include numerical diffusion in the $z$-direction.

Following the approach of \cite{frei2023fullf} and \cite{mencke2025extended}, we use Bohm fluid boundary conditions for $U_{\|}$ and $J_{\|}$ \cite{Loizu2012,Mosetto2015} at the sheath entrances
\begin{subequations} \label{eq:vparbc}
\begin{align} 
U_{\|}\left(z=\pm L_{z}/2\right)&=\pm c_s & \nonumber \\
& = \pm \sqrt{T_{eDK} + \tau_i T_{iDK} },\\
J_{\|}\left(z=\pm L_{z}/2\right)&=\pm  N_{DK}\left(c_s-\sqrt{T_{eDK}} e^{   \Lambda  - \phi_{DK} / T_{eDK}}\right), & \nonumber \\
\end{align}
\end{subequations}
while homogeneous Neumann boundary conditions are used for the remaining boundaries for $J_{\|}$ and $U_{\|}$. For all other DK fields, homogeneous Neumann boundary conditions are applied at all boundaries while we apply homogeneous Neumann boundary conditions in the parallel direction for the GK fields and periodic boundary conditions in the perpendicular plane.

The plasma sources are chosen as
\begin{subequations}\label{eq:sources}
\begin{align}
    S_{i}&=A_{N}\left(\bm x\right) F_{Mi}+A_{E}\left(\bm x\right) \left(s_{\| i}^2+x_i-\frac{3}{2}\right)N_{DK}F_{Mi},
\intertext{where $\left\langle \gyroavg{A_{N}\left(\bm x\right)} \right\rangle_{DK}=A_N$ is assumed and analogously for $A_E$. Therefore, we order $S_i$ on the DK timescale ($S_i\sim \partial_t F_{iDK}$). The projection in the DK hierarchy, \eqref{eq:moment_hierachy_DK}, is}
    S^{pj}_{i0}&=A_N\left(\bm R\right) \delta^{pj}_{00}+\frac{\delta_{20}^{pj}}{\sqrt{2}} N_{DK}A_E\left(\bm R\right)-\delta_{01}^{pj}N_{DK}A_E\left(\bm R\right),\\
\intertext{and the projections in the GK hierarchy, \eqref{eq:moment_hierachy_gk_i}, is given by}
    \delta S^{pj}_{i1}&=\delta^{p}_{0}\left(K_{j}\left(b_i\right)-\delta^{j}_0\right)A_N+\frac{\delta^{p}_{2}}{\sqrt{2}}N_{DK}\left(K_{j}\left(b_i\right)-\delta^{j}_0\right)A_E\nonumber\\
    &+\delta^{p}_{0}N_{DK}\left(2j K_{j}\left(b_i\right)-\left(j+1\right)K_{j+1}\left(b_i\right)-jK_{j-1}\left(b_i\right)+\delta^{j}_1\right)A_E.\label{eq:Source_GK}\\
\intertext{For the electrons in Eqs. (\ref{eq:dt_Ne}) and (\ref{eq:dt_Te}), the sources are}
    S_{N}&=A_{N}\left(\bm R\right),\quad S_{Te}=A_{Te}\left(\bm R\right),
    \intertext{with}
    K_n\left(x\right)&= \frac{1}{n!}\left(\frac{x}{2}\right)^{2n}e^{-\left(x/2\right)^2},\quad \text{and}\quad b_i=\frac{v_{Th i}k_{\perp}}{\Omega_i}.
\end{align}   
\end{subequations}
Finally, we do not include a momentum source such that $S_U=0$ in \eqref{eq:Upar}.

Simulations that solve the full system, including DK and GK fluctuations, are referred to as full simulations. For comparison, simulations are run where $\N^{pj}_{iGK}=0$ and $\phi_{GK}=0$. These are referred to as DK simulations.

We start the simulations from smooth analytical profiles satisfying the boundary conditions in \eqref{eq:vparbc} and evolve the DK simulations at first. Quasi-steady state is achieved after approximately $165 $ $c_{s0} / R$ time units using a small number of gyromoments ($P_{DK}=2$ and $J_{DK}=1$). Afterwards, the number of moments, $P_{DK}$ and $J_{DK}$, is increased and the evolution of $\phi_{GK}$ and $\N^{pj}_{iGK}$ is included. We choose $P_{DK}=2J_{DK}$ for the DK moments and $P_{GK}\lesssim J_{GK}$ for the GK moments to study the influence of FLR effects on the simulations. When a quasi-steady state is reached, the particles and energy injected by the sources are transported slowly across the field lines by turbulent transport and rapidly along the field lines to the sheaths where the plasma outflow at the sheath entrances balances the sources. All analyses are carried out during this quasi-steady state.


For our simulations, we consider a helium LAPD plasma \cite{Gekelman1991} with parameters as in Refs. \cite{Rogers2010,frei2023fullf,mencke2025extended}. The simulation parameters are: $n_{e0} = 2 \times 10^{12}\: \mathrm{cm}^{-3}$, $T_{e0} = 6\: \mathrm{eV}$, $T_{i0} = 3 \: \mathrm{eV}$, $\Omega_{i} \sim 960\:\mathrm{kHz}$, $\rho_{s0} = 1.4\:\mathrm{cm}$, $c_{s0}= 1.3 \times 10^{6}\:\mathrm{cm/s}$ , $m_i / m_e =  400$, $\nu_0 = 0.03$, and $\tau_{i}=0.5$. The LAPD vacuum chamber has a radius $R \simeq 0.56\: \mathrm{m}$, therefore we assume $R = 40 \rho_{s0}$ and, with a parallel extension $L_z \simeq 18\: \mathrm{m}$, we impose $L_z=36 R$. With these parameters, the reference time is $R / c_{s0}  \sim 43\:\mu\mathrm{s}$. We set the perpendicular extension of the domain $L_x=L_y=100\rho_{s0}$, sufficiently large for the turbulent structures not to be affected by the perpendicular boundary conditions. We choose our sources in Eq. (\ref{eq:sources}) to have radial extension from the axis $r_s$ and radial gradient length $L_s$, that is
\begin{align} \label{eq:ANxy}
     {A}_N (x,y)  = \frac{\mathcal{A}_{N0}}{2}    \left[ 1 - \tanh\left( \frac{r -r_s}{L_s}\right)  \right] + \mathcal{A}_{N \infty},
\end{align}
where $r=\sqrt{x^2+y^2}$ and equivalent expressions are used for ${A}_{E}$, $A_{T_e}$, and $\mathcal{A}_N$ with $\mathcal{A}_{N0},\mathcal{A}_{E0}$, and $\mathcal{A}_{Te0}$ representing the bulk plasma sources. The constant terms, $\mathcal{A}_{N\infty},\mathcal{A}_{E\infty},\mathcal{A}_{Te \infty}\ll  \mathcal{A}_{N0},\mathcal{A}_{Ti0},\mathcal{A}_{Te0}$, avoid negative values of temperature and density and are chosen to have small amplitude. The source parameters are chosen as $\mathcal{A}_{N\infty} = \mathcal{A}_{T_e \infty}  =  0.004$, $\mathcal{A}_{E\infty} = 0.001$, $\mathcal{A}_{N0}=\mathcal{A}_{Te0}=\mathcal{A}_{E0}  = 0.04$, $L_s = 1  \rho_{s0}$, $r_s = 20 \rho_{s0}$. 

The perpendicular grid resolution is $N_x = N_y = 192$, which corresponds to a grid spacing of $\sim 0.52 \rho_{s0}$. The number of parallel grid points is $N_z=64$, which corresponds to a grid spacing of $\sim 0.56 R$. For the DK fields ($n_e$, $T_e$, $U_{\|}$, $\Omega$, $J_{\|}$, and $\N^{pj}_{iDK}$) diffusion parameters are chosen as $\eta_{zDK}\simeq 4$ and $\eta_{\perp}\simeq 0.5$. For the Fourier grid, we impose $N_{kx}=N_{ky}=96$ such that $\max \left(k_x\right)=\max \left(k_y\right)=3.08/\rho_{s0}$. For $\N^{pj}_{iGK}$, $\eta_{z GK} \simeq 4$ is the same value as for the DK fields. We note that no numerical diffusion for the GK fields is required since the $\nabla_\perp^2 \N^{pj}_{iGK}$ term in \eqref{eq:C_iGK} provides sufficiently large damping for high $k_{\perp}$ modes and also ensures strong damping for modes with $k_{\perp}\gtrsim 1/\rho_{s0}$. Finally, we use $\eta_{GK}=10^{-7}$.

\section{Simulation results}\label{sec:sim_res}
In this section, we present simulations based on the model described in Secs. \ref{sec:DK_sys}, \ref{sec:GK_sys}, and \ref{sec:field_equations}, using the parameters of the LAPD experiment described in \secref{sec:Num_impl}. We show the equilibrium DK and GK gyromoments in \secref{subsec:eq_gm}. In \secref{subsec:conv} we demonstrate spectral convergence of our simulations by comparing the ion distribution function with a local bi-Maxwellian and by investigating the effect of the number of gyromoments on the statistical properties of $\phi_{DK}$. We investigate the turbulence properties of the DK fields in \secref{subsec:DK_prop}. Finally, in \secref{subsec:prob_GK} we analyze the properties of the GK fields.

\subsection{Equilibrium gyromoments}\label{subsec:eq_gm}
\begin{figure}
    \centering
    \includegraphics[width=\linewidth]{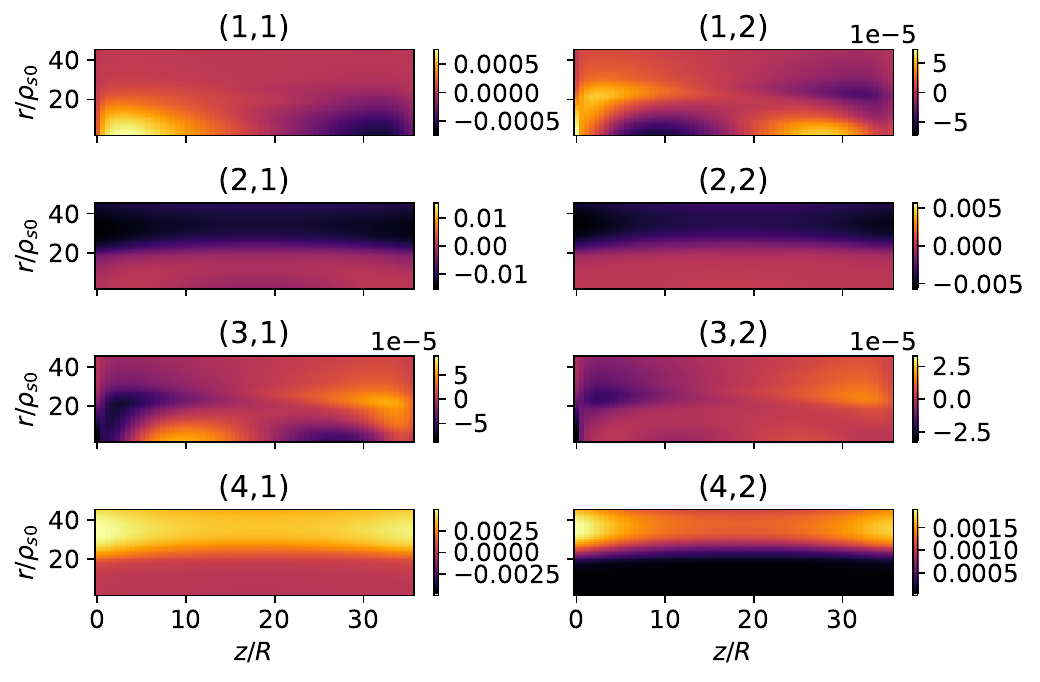}
    \caption{The spatial and azimuthal average of the DK moments, $\left\langle \N^{pj}_{iDK} \right\rangle_{t,\vartheta}$, for a full simulation with $\left(P_{DK},J_{DK},P_{GK},J_{GK}\right)=\left(6,3,6,5\right)$ for $p=1-4$ (rows) and $j=1,2$ (columns). Moments with odd $p$ are antisymmetric around $z=L_z/2$ and moments with even $p$ are symmetric around $z=L_z/2$.}
    \label{fig:npj_equil}
\end{figure}
\begin{figure}
    \centering
    \includegraphics[width=\linewidth]{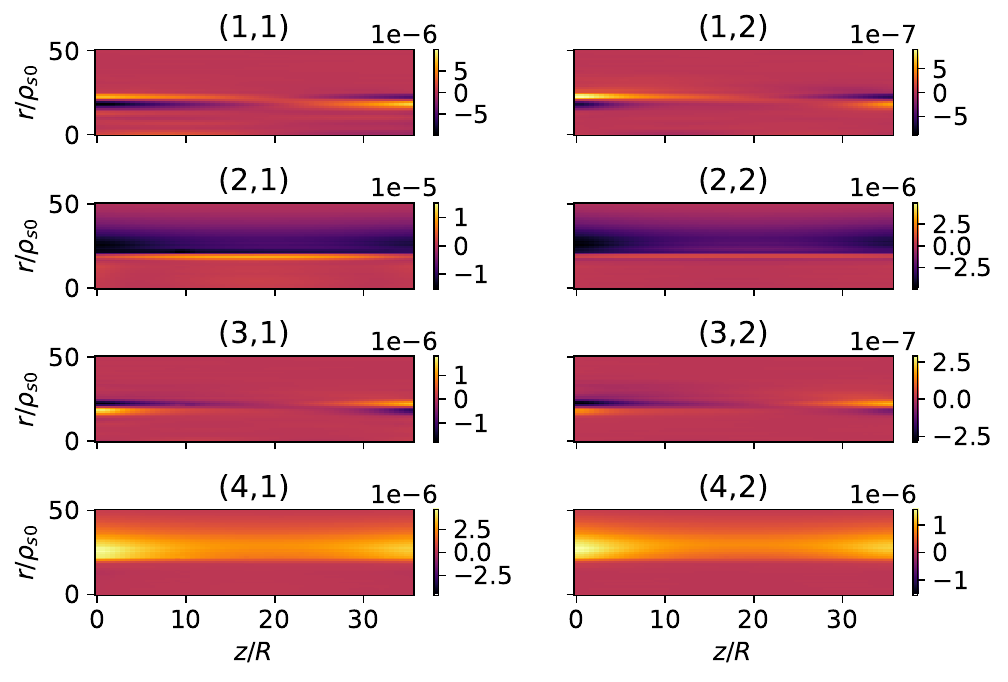}
    \caption{The spatial and azimuthal average of $\N^{pj}_{iGK}$, $\left\langle \N^{pj}_{iGK} \right\rangle_{t,\vartheta}$, for a full simulation with $\left(P_{DK},J_{DK},P_{GK},J_{GK}\right)=\left(6,3,6,5\right)$ for $p=1-4$ (rows) and $j=1,2$ (columns). Moments with odd $p$ are antisymetric around $z=L_z/2$ and moments with even $p$ are symmetric around $z=L_z/2$.}
    \label{fig:npj_gk_equil}
\end{figure}

The spatial and azimuthal average, defined as 
\begin{equation}
    \left\langle \cdot \right\rangle_{t,\vartheta}=\frac{1}{2\pi}\int_0^{2\pi} d\vartheta\frac{1}{\Delta t}\int_{t_0}^{t_0+\Delta t}d t \left(\cdot\right),\label{eq:avg}
\end{equation}
of the DK gyromoments, $\left\langle \N_{iDK}^{pj} \right\rangle_{t,\vartheta}$, with $\vartheta=\arctan y/x$ are shown in \figref{fig:npj_equil}. The gyromoments with odd (even) $p$ are antisymmetric (symmetric) around $z=L_z/2$. Furthermore, the amplitude of the moments decreases with $p$ and $j$. We note that the plasma equilibrium is very similar to previous gyromoment simulations of an LAPD plasma \cite{mencke2025extended}.

Figure \ref{fig:npj_gk_equil} shows the spatial and azimuthal average of the GK moments, $\left\langle \N_{iGK}^{pj} \right\rangle_{t,\vartheta}$. Similar symmetry properties as for the DK moments are observed, although the GK moments have significantly smaller amplitude. Furthermore, the amplitude of $\left\langle \N_{iGK}^{pj} \right\rangle_{t,\vartheta}$ is localized around the source gradient $r\sim 20\rho_{s0}$ since $\delta S_{i1}^{pj}$ is the largest in that region.

\begin{figure}
    \centering
    \includegraphics[width=0.65\linewidth]{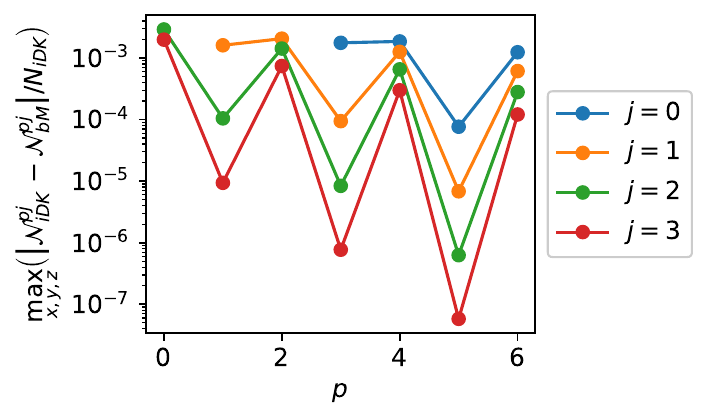}
    \caption{The spatial maximum of the difference between the DK moments and the moments of a bi-Maxwellian calculated using \eqref{eq:bi_Max} with the local parallel and perpendicular temperatures normalized to the local density for $\left(P_{DK},J_{DK},P_{GK},J_{GK}\right)=\left(6,3,6,5\right)$. Small deviations from a bi-Maxwellian are observed.}
    \label{fig:Npj_Max_compare}
\end{figure}

In order to quantify the kinetic part of the distribution function, we compare the moments from the full simulations with those of a local bi-Maxwellian \cite{mencke2025extended}
\begin{subequations}
\begin{align}
    F_{bM}&=\frac{N_im_i^{3/2}}{\left(2\pi\right)^{3/2}\sqrt{T_{\|}}T_{\perp i}}\exp\left(-\frac{m_i\left(v_{\|}-U_{\| i}\right)^2}{T_{\| i}}-\frac{\mu B}{T_{\perp i}}\right),
\intertext{with the projection on the Hermite-Laguerre basis in \eqref{eq:DK_Npj_exp} given by}
    \N^{pj}_{bM}&=\begin{cases}
        N_i\frac{\left(-1\right)^{j}\sqrt{p!}}{\sqrt{2^{p}}\left(p/2\right)!}\left(\frac{T_{\| i}}{T_{i0}}-1\right)^{p/2}\left(\frac{T_{\perp i}}{T_{i0}}-1\right)^{j}& \text{for even}\: p,\\
        0 & \text{for odd}\: p.
    \end{cases}\label{eq:bi_Max}
\end{align}
\end{subequations}
We use the local $T_{\perp i}$, $T_{\| i}$, and $N_{i}=N_{DK}$ in \eqref{eq:bi_Max} and show the spatially maximal deviation for the gyromoments in the full-f simulation for each $p$ and $j$ in \figref{fig:Npj_Max_compare}. Small deviations are observed due to the high collisionality in the simulations, similarly to DK-ordered gyromoment simulations of the same system \cite{mencke2025extended}.

\subsection{Convergence study}\label{subsec:conv}

\begin{figure}
    \centering
    \includegraphics[width=1\linewidth]{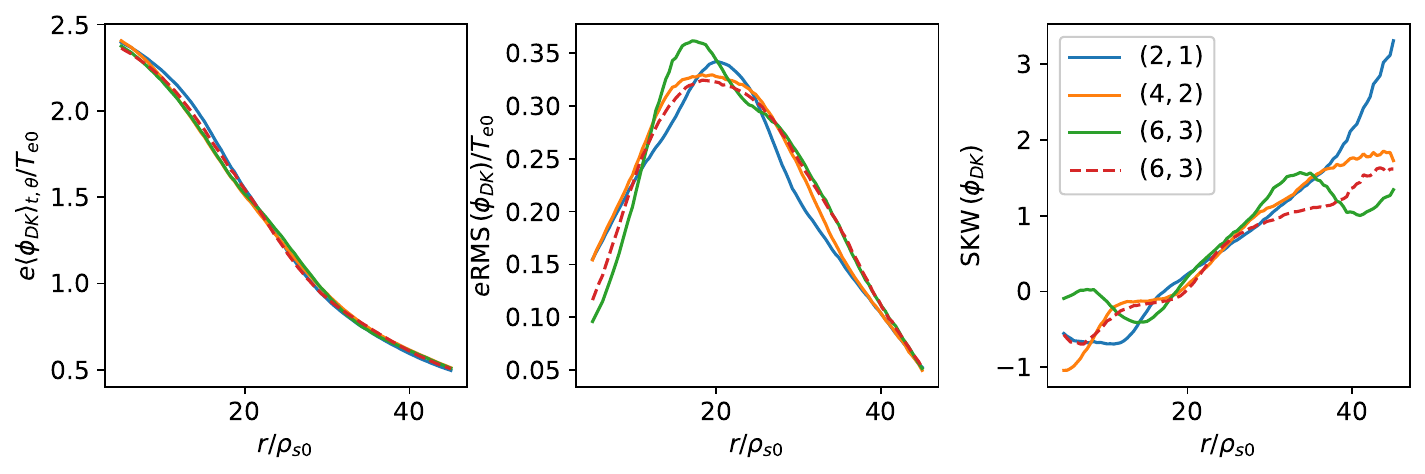}
    \caption{The azimuthal and temporal average, root mean square, and skewness of $\phi_{DK}$ at $z=L_z/2$ for different choices of $\left(P_{DK},J_{DK},P_{GK},J_{GK}\right)$. For the GK simulations, $P_{DK}=P_{GK}$ and $J_{GK}=5$. The full (solid) and DK (dashed) models are compared. The simulations are considered converged at $\left(P_{DK},J_{DK}\right)=\left(2,1\right)$. The presence of GK fields does not impact $\phi_{DK}$ noticeably.}
    \label{fig:phi_conv}
\end{figure}

Since deviations from a bi-Maxwellian are small, a small number of moments is needed to reach convergence. We investigate the statistical properties of the DK electrostatic potential, $\phi_{DK}$, and how these are affected by the number of evolved moments  in \figref{fig:phi_conv}. We compare the temporal and azimuthal average of $\phi_{DK}$, defined in \eqref{eq:avg}, the root mean square,
\begin{subequations}\label{eq:avg_RMS_SKW}
\begin{align}
    RMS\left(\phi_{DK}\right)&=\sqrt{\left\langle \left(\left(\phi_{DK}\right)-\left\langle \phi_{DK}\right\rangle_{t,\vartheta}\right)^2\right\rangle_{t,\vartheta} },
\intertext{and the skewness}
    SKW\left(\phi_{DK}\right)&=\left\langle \left(\left(\phi_{DK}\right)-\left\langle \phi_{DK}\right\rangle_{t,\vartheta}\right)^3\right\rangle_{t,\vartheta}/RMS^3\left(\phi_{DK}\right),
\end{align}
\end{subequations}
for different choices of $P_{DK},J_{DK},P_{GK},J_{GK}$ and for the full and DK simulations. It is observed that spectral convergence is obtained at $\left(P_{DK},J_{DK}\right)=\left(2,1\right)$ for both the full and DK simulations and the average, RMS, and skewness of $\phi_{DK}$ are not significantly affected by the presence of GK fields. 
A similar trend is observed for all other DK fields. Both models and all choices of $\left(P_{DK},J_{DK}\right)$ display a negative skewness at $r\lesssim 20\rho_{s0}$, associated with the propagation of holes, and a positive skewness for $r\gtrsim 20\rho_{s0}$, associated with blobs, similar to simulations of the tokamak boundary. Similar statistical behavior of the fields was reported in previous simulations of LAPD plasmas \cite{frei2023fullf,mencke2025extended}. In the following, we analyse the results of the simulations with $\left(P_{DK},J_{DK}\right)=\left(6,3\right)$.

\subsection{Properties of the DK fields}\label{subsec:DK_prop}

\begin{figure}
    \centering
    \includegraphics[width=1\linewidth]{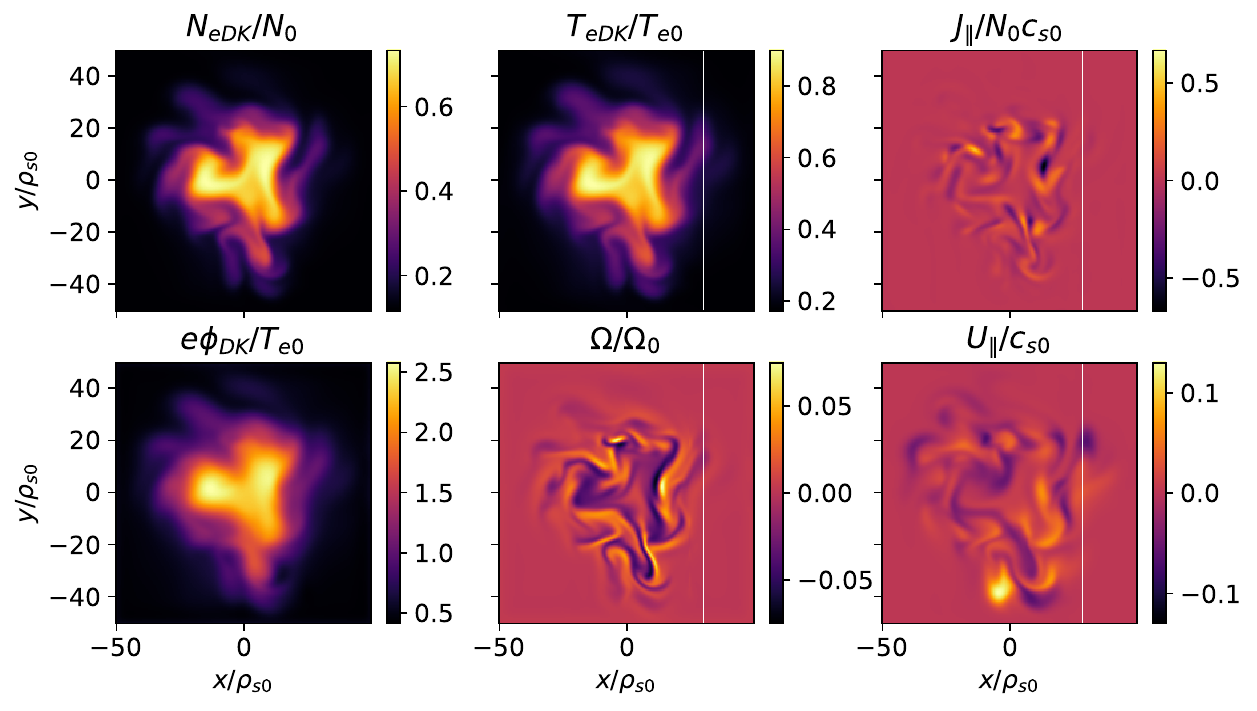}
    \caption{Perpendicular, $\left(x,y\right)$, snapshot of $N_{DK}$ (top left), $T_{eDK}$ (top center), $J_{\|}$ (top right), $\phi_{DK}$ (bottom left), $\Omega$ (bottom center), and $U_{\|}$ (bottom right) for a full simulation with $\left(P_{DK},J_{DK},P_{GK},J_{GK}\right)=\left(6,3,6,5\right)$ at $z=L_z/2$. Turbulent structures are observed at the source-gradient region ($r\sim r_s\sim 20\rho_{s0}$).}
    \label{fig:snapshot_dk_perp}
\end{figure}

\begin{figure}
    \centering
    \includegraphics[width=1\linewidth]{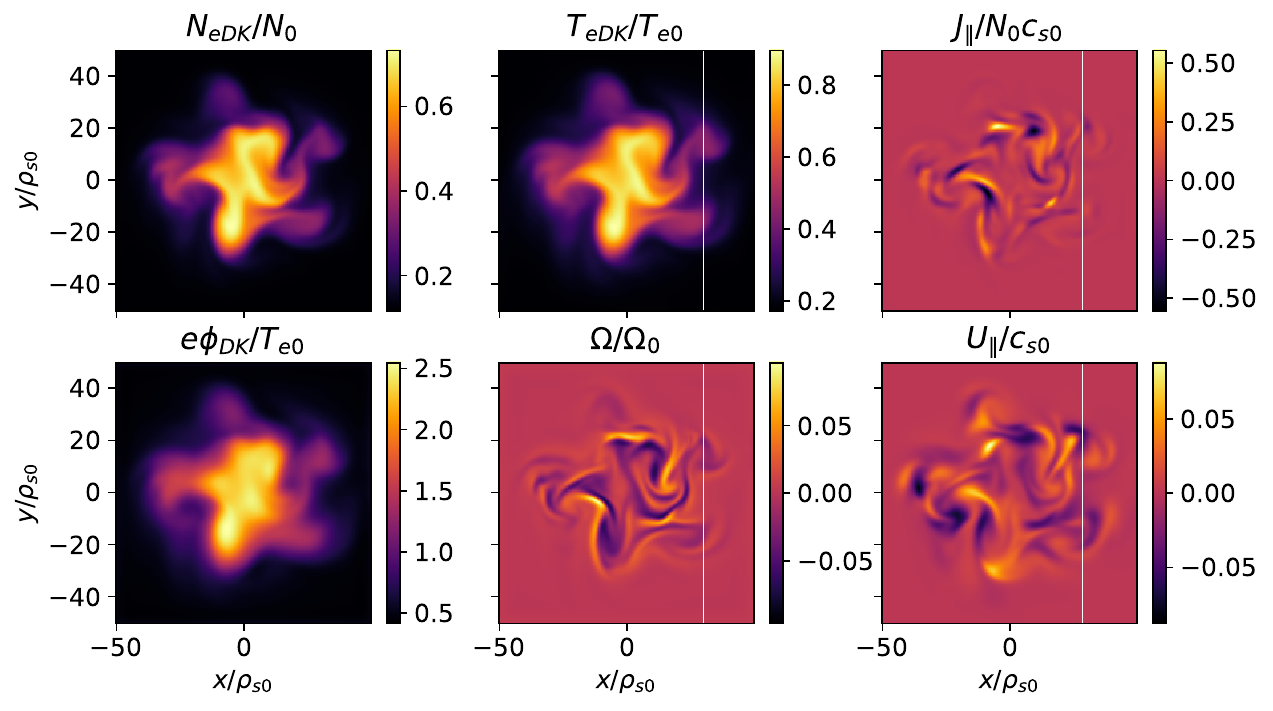}
    \caption{Same as \figref{fig:snapshot_dk_perp} for a DK simulation with $\left(P_{DK},J_{DK}\right)=\left(6,3\right)$. Similar structures as for the full model in \figref{fig:snapshot_dk_perp} are observed. }
    \label{fig:snapshot_dk_perp_dk}
\end{figure}

\begin{figure}
    \centering
    \includegraphics[width=1\linewidth]{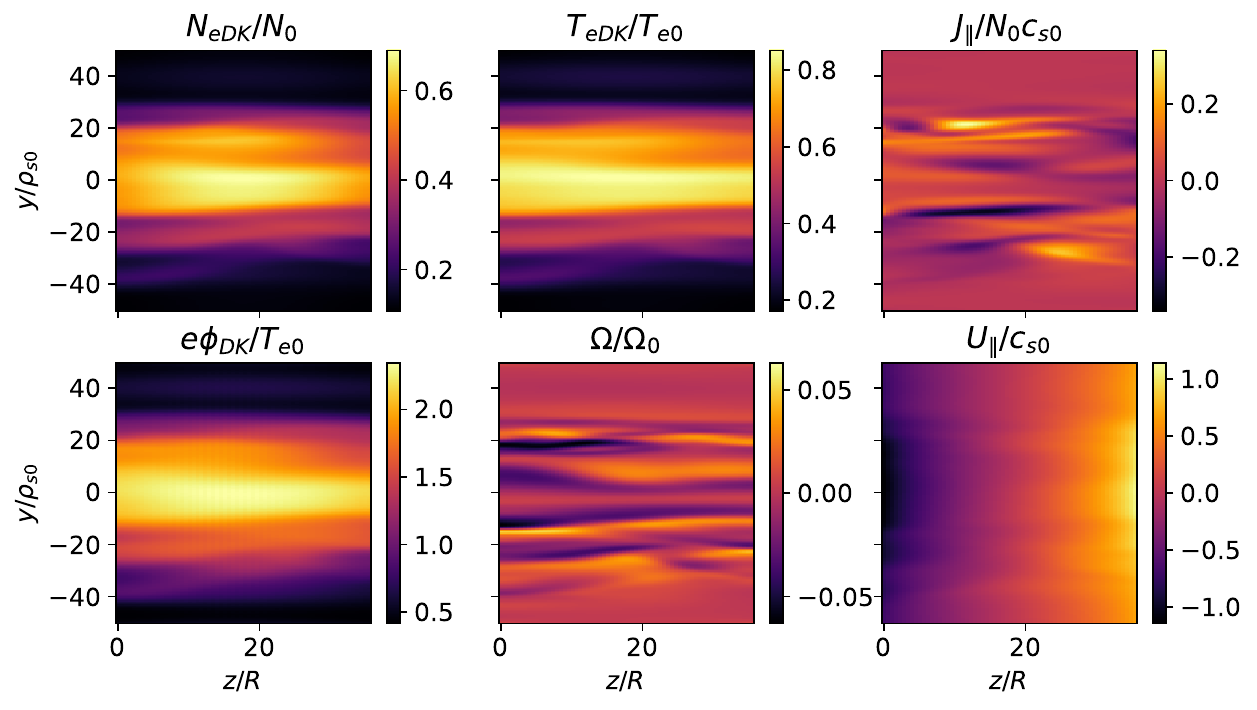}
    \caption{Parallel, $\left(z,y\right)$, snapshot of $N_{DK}$ (top left), $T_{eDK}$ (top center), $J_{\|}$ (top right), $\phi_{DK}$ (bottom left), $\Omega$ (bottom center), and $U_{\|}$ (bottom right) for a full simulation with $\left(P_{DK},J_{DK},P_{GK},J_{GK}\right)=\left(6,3,6,5\right)$ at $x=0$. The turbulent structures are very elongated along $z$.}
    \label{fig:snapshot_dk_par}
\end{figure}

We consider the turbulent behavior of the DK fields. Figure \ref{fig:snapshot_dk_perp} shows a perpendicular snapshot of the DK fields for the full simulation, while \figref{fig:snapshot_dk_perp_dk} shows the same snapshot for the DK simulation. We note that for both simulations, all fields display large turbulent structures in the perpendicular plane. The DK and full simulations are very similar, indicating that the inclusion of the GK fields do not significantly affect the system. A snapshot of the DK fields in the parallel plane is shown in \figref{fig:snapshot_dk_par} for the full simulations. We observe elongated structures in the $z$-direction. Similar properties are observed for the DK simulations (not shown here). 


\begin{figure}
    \centering
    \includegraphics[width=\linewidth]{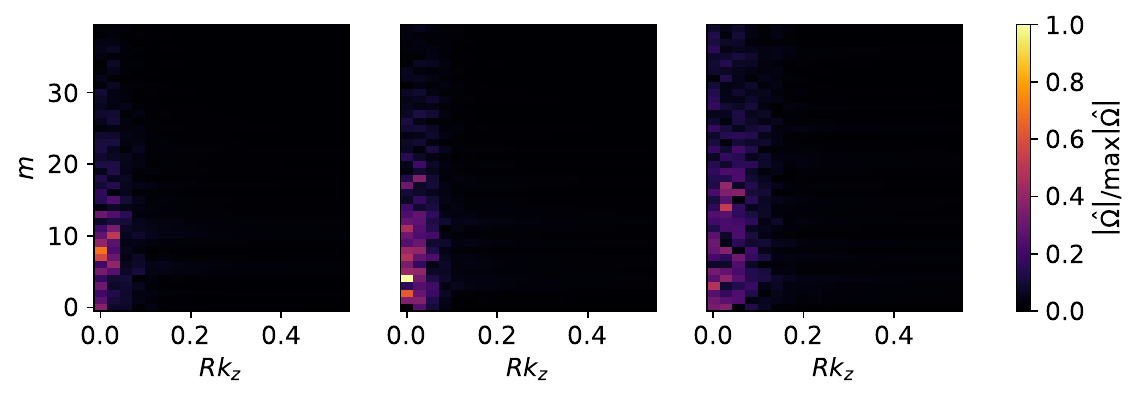}
    \caption{Power spectrum of $\Omega=\sum_{m,\omega}\hat{\Omega}e^{i\left(m\vartheta+k_z z\right)}$ at the radial distance, $r=\sqrt{x^2+y^2}=20\rho_{s0}$ for a full simulation with $\left(P_{DK},J_{DK},P_{GK},J_{GK}\right)=\left(6,3,6,5\right)$ (left), the DK simulation with $\left(P_{DK},J_{DK}\right)=\left(6,3\right)$ (center), and a full simulation with $\nu_{i0}$ in \eqref{eq:C_iGK} reduced by a factor of $1000$ and with $A_E$ and $A_N$ increased by a factor $10$ in \eqref{eq:Source_GK} (right).}
    \label{fig:Power_m_kz}
\end{figure}

Figure \ref{fig:Power_m_kz} compares the power spectrum of $\Omega$ as a function of the parallel wave number, $k_z$, and azimuthal mode number, $m$. We observe low-$m$ modes elongated in $z$. The full and DK simulations have similar power spectrum. To test the interaction between DK and GK fields, we artificially lower the collisionality in \eqref{eq:C_iGK} by a factor of $1000$, setting  $\nu_{i0}\sim 0.00234 c_{s0}/R$, and we increase $A_E$ and $A_N$ by a factor $10$ in \eqref{eq:Source_GK}. By reducing the GK collisional term and increasing the GK source term, we observe an increase in high $m$ modes indicating that in a system with lower collisionality and larger sources, the GK fields can introduce high $k_\perp$-modes in the DK fields.

\subsection{Properties of gyrokinetic fields}\label{subsec:prob_GK}
\begin{figure}
    \centering
    \includegraphics[width=\linewidth]{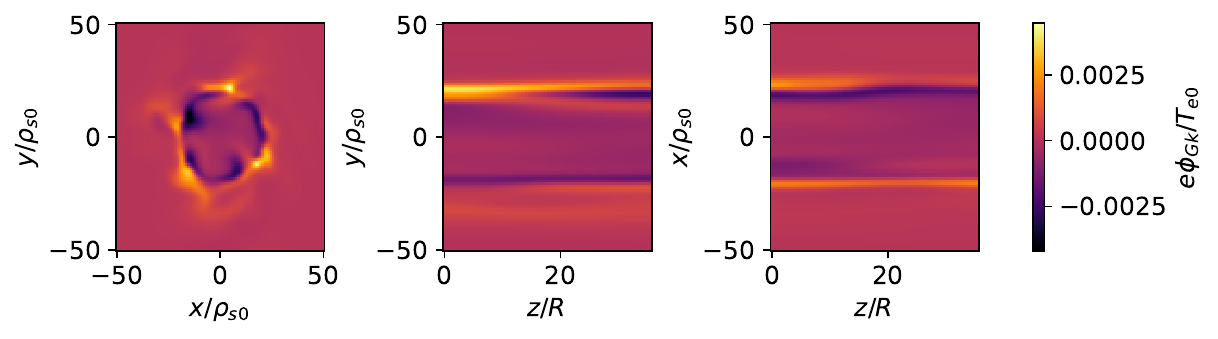}
    \caption{Snapshots of $\phi_{GK}$ in real space at $z=L_z/2$ (left), $x=0$ (middle), and $y=0$ (right) for $\left(P_{DK},J_{DK},P_{GK},J_{GK}\right)=\left(6,3,6,5\right)$. We note that $\phi_{GK}$ has small amplitude with small scale structures in the perpendicular plane and is almost constant along $z$.}
    \label{fig:phi_gk_real}
\end{figure}

\begin{figure}
    \centering
    \includegraphics[width=\linewidth]{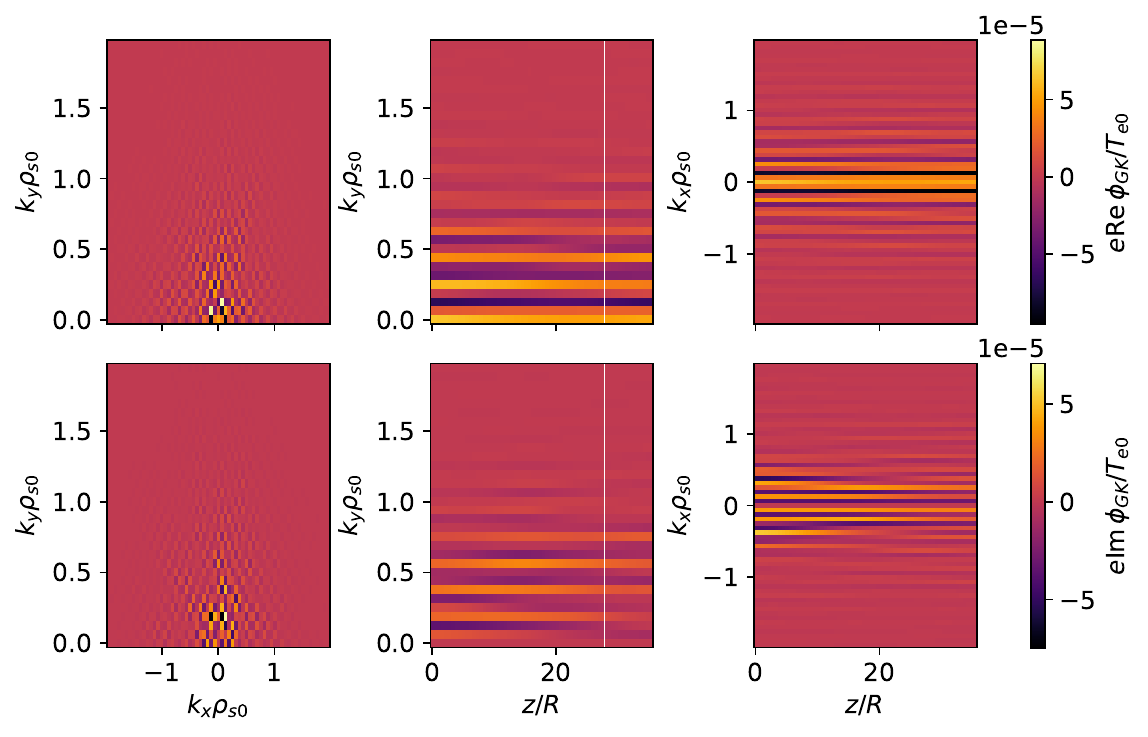}
    \caption{Snapshots of $\phi_{GK}$ in Fourier $\left(k_x,k_y,z\right)$ space at $z=0$ (left), $k_x=0$ (middle), and $k_y=0$ (right) real part (top) and imaginary part (bottom) for $\left(P_{DK},J_{DK},P_{GK},J_{GK}\right)=\left(6,3,6,5\right)$.}
    \label{fig:phi_gk_complx}
\end{figure}

In \figref{fig:phi_gk_real}, $\phi_{GK}$ is shown in real space $\left(x,y,z\right)$ while \figref{fig:phi_gk_complx} shows $\phi_{GK}$ in $\left(k_x,k_y,z\right)$ space. The GK potential, $\phi_{GK}$, depends weakly on the $z$-direction and is dominated by $k_\perp \rho_s\lesssim 1$ modes. Comparing with \figref{fig:snapshot_dk_perp}, it is seen that $\phi_{GK}$ is one order of magnitude smaller than $\phi_{DK}$ and from \figref{fig:phi_gk_complx}, $k_\perp\rho_{s0}\sim 1$ modes are observed to exist even though they are not dominant. This is consistent with the ordering imposed in \secref{sec:oredering}. Due to the high collisionality, the spatial diffusion induced by the $\nabla_{\perp}^2\N^{pj}_{iGK}$ term in \eqref{eq:C_iGK} suppresses high $k_{\perp}$ modes for $\N^{pj}_{iGK}$. Since in \eqref{eq:Poisson_GK}, $\rho_i^*\propto \exp\left(-2T_{i0}k_{\perp}^2/m_i\Omega_i^2\right)$ [see \eqref{eq:varrho_i}], we further dampen high $k_{\perp}$ modes in $\phi_{GK}$.


\begin{figure}
    \centering
    \includegraphics[width=0.6\linewidth]{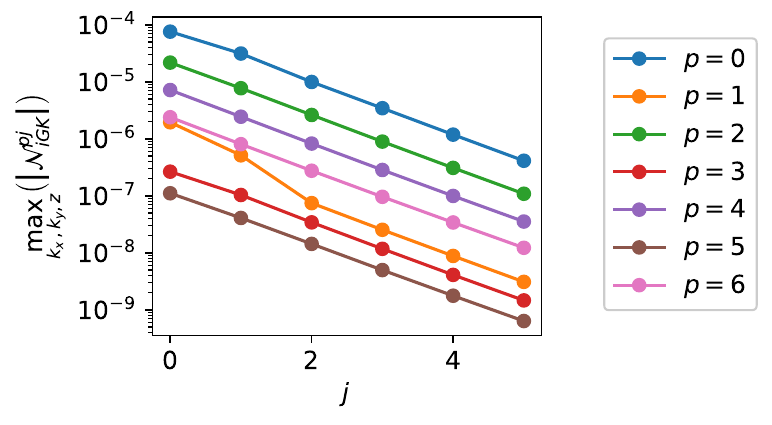}
    \caption{Spatial maximum of the amplitude of the GK moments, $\max_{k_x,k_y,z}\left( \left|\N^{pj}_{iGK}\right| \right)$, for $\left(P_{DK},J_{DK},P_{GK},J_{GK}\right)=\left(6,3,6,5\right)$. The magnitude quickly decreases with $p$ and $j$.}
    \label{fig:npj_gk_max}
\end{figure}

Finally, the amplitude of the GK moments are shown in \figref{fig:npj_gk_max}. The amplitude of the moments, $\N^{pj}_{GK}$, rapidly decreases with $p$ and $j$. This is due to the large collisionality which suppresses high $p$ and $j$ moments, as seen from \eqref{eq:C_iGK}.

\section{Linear analysis of the turbulent modes
}\label{sec:linear}
For studying the turbulent behavior of the simulations observed in \figref{fig:Power_m_kz}, the system of Eqs. (\ref{eq:dt_Ne}), (\ref{eq:dt_Te}), (\ref{eq:moment_hierachy_i}), (\ref{eq:moment_hierachy_gk_i}), (\ref{eq:Upar}), (\ref{eq:dt_Omega}), (\ref{eq:Omega_impl}), (\ref{eq:dt_J}),  and (\ref{eq:Poisson_GK}) is linearized. Specifically, each DK field, $f_{DK}$, is split as
\begin{equation}
    f_{DK}=\Bar{f}_{DK}\left(r\right)+\Tilde{f}_{DK}\left(r\right)e^{i\left(m\vartheta+k_{\|}z -\omega t\right)},
\end{equation}
where the equilibrium, $\Bar{f}_{DK}$, is calculated from the full 3D turbulent simulations by performing a temporal and azimuthal averaging at $z=L_z/2$. For the GK field, $f_{GK}$, we apply a local approximation,
\begin{equation}
    f_{GK}=\Bar{f}_{GK}\left(r\right)+\Tilde{f}_{GK}\delta\left(r-r_{g}\right)e^{i\left(m\vartheta+k_{\|}z -\omega t\right)},
\end{equation}
where $r_g$ is the radial position at which the GK fluctuations are localized. To calculate $\Bar{f}_{GK}$ we first perform an inverse Fourier transform in the perpendicular plane of the fields in the global full 3D simulations. Then, the temporal and azimuthal averaging at $z=L_z/2$ are evaluated.

\begin{figure}
    \centering
    \includegraphics[width=\linewidth]{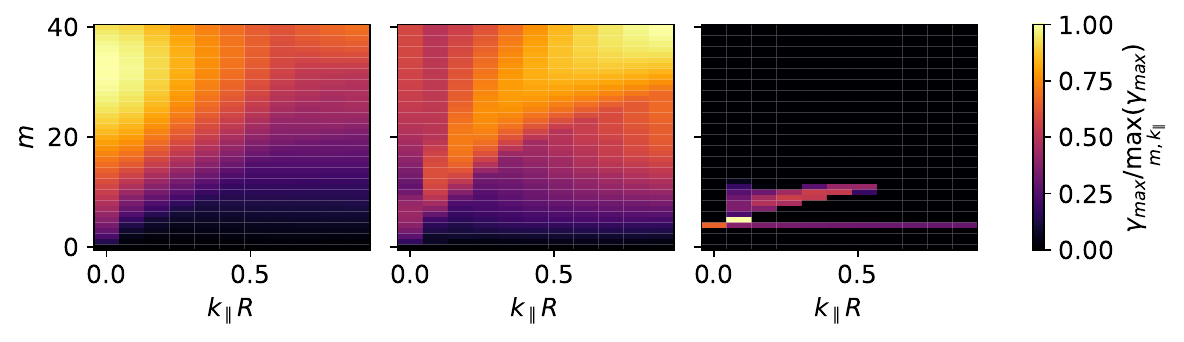}
    \caption{The linear growth rates of the system of Eqs. (\ref{eq:dt_Ne}), (\ref{eq:dt_Te}) (\ref{eq:dt_Omega}), (\ref{eq:Omega_impl}), (\ref{eq:dt_J}), (\ref{eq:Upar}),  and (\ref{eq:Poisson_GK}) with $\left(P_{DK},J_{DK},P_{GK},J_{GK}\right)=\left(4,2,4,3\right)$, for the full system (left), for the DK system with $f_{GK}=0$ (center), and for the GK system with $\Tilde{f}_{DK}=0$ (right), $r_g=22\rho_{s0}$. The simulations with physical GK collisionality and sources are used to calculate the equilibrium fields for the DK system while the simulations with $\nu_{i0}$ in \eqref{eq:C_iGK} reduced by a factor of $1000$ and with $A_E$ and $A_N$ increased by a factor $10$ in \eqref{eq:Source_GK} are used for the two other systems. The maximum growth rates for the full model is $\max_{m,k_{\|}}\left(\gamma_{max}\right)=1.62 c_{s0}/R$, for the DK model it is $\max_{m,k_{\|}}\left(\gamma_{max}\right)=0.91 c_{s0}/R$ and for the GK model it is $\max_{m,k_{\|}}\left(\gamma_{max}\right)=0.014 c_{s0}/R$. The highest growth rate for the full system is observed at $m=33$ and $k_{\|} =0 $. }
    \label{fig:linear}
\end{figure}

In \figref{fig:linear} we show the dominant linear growth-rate, $\gamma\equiv\Im\omega$, as function of $k_\|$ and $m$ for the full model, the DK model (where we impose $\N^{pj}_{iGK}=0$ and $\phi_{GK}=0$), and the GK model (where we impose $\Tilde{f}_{DK}=0$). We only consider $m\leq 40$ since the numerical diffusion for the DK field significantly dampens modes with higher $m$. The equilibrium fields, $\Bar{f}$, for the DK model are calculated using the simulations with physical collisionality and sources. With the physical parameters, the full model yields linear growth rates  identical to that of the DK model while the GK model is stable for all choices of $m$ and $k_{\|}$.  For this reason, for the full and GK model, we calculate the equilibrium fields using the simulations where $\nu_{i0}$ in \eqref{eq:C_iGK} is reduced by a factor of $1000$ and where $A_E$ and $A_N$ are increased by a factor $10$ in \eqref{eq:Source_GK}. Furthermore, we neglect $\C^{pj}_{ii1}$ in \eqref{eq:moment_hierachy_gk_i}.

By artificially suppressing different terms, we study the linear drive of the dominant instabilities. For the DK model, the peak growth rate is a Kelvin-Helmholz instability \cite{d1965kelvin}. More specifically, the modes are driven by the $\poissonbracket{\phi_{DK}}{\Omega}$ and the $\poissonbracket{\nabla_{\perp} \phi_{DK}\cdot}{\boldsymbol{\omega}}$ terms in \eqref{eq:dt_Omega}. This finding is consistent with previous long-wavelength simulations of linear plasma devices \cite{frei2023fullf,mencke2025extended,Rogers2010}. 

For the GK model, the dominant instability is at $k_{\|}=0.087 /R$ and $m=5$ and it is driven by the  $\poissonbracket{\phi_{DK}}{\N^{pj}_{iGK}}$ term in \eqref{eq:moment_hierachy_gk_i} for $p=j=0$. It is driven by shear flow and a density gradient since both the GK density gradient and the gradient of the DK electrostatic potential are necessary for the instability to appear. We note that for $m=5$, using \eqref{eq:C_iGK}, $\C^{00}_{ii1}\sim -0.051 \nu_{i0}\tau_i\N^{00}_{iGK} $. Therefore, collisions damp this instability with $\gamma\sim 0.014 c_{s0}/R$ when using the physical collisionality.



The dominant instability for the full model is a Kelvin-Helmholz mode with the growth rate enhanced by the $\poissonbracket{\phi_{DK}}{\N^{pj}_{iGK}}$ term in \eqref{eq:moment_hierachy_gk_i} for $p=j=0$ and the $\poissonbracket{\phi_{GK}}{\N^{pj}_{iDK}}$ term in \eqref{eq:moment_hierachy_DK} for $p=j=0$. Through a DK density gradient, the Kelvin-Helmholz mode is coupled to the GK fields. We note that $\Bar{\phi}_{GK}$ is an order of magnitude smaller than the amplitude of $\Bar{\phi}_{GK}$. Therefore, we do not observe a significant influence of $\Bar{\phi}_{GK}$ in the 1D linear simulations.

In order to study the development of small-scale structures observed for the low-collisionality and larger-GK-sources simulations observed in \figref{fig:Power_m_kz}, we assume that $k_\|=0$, $F_{iDK}$ is a Maxwellian constant in time and space, and $F_{iGK}$ is an isotropic Maxwellian. Focusing on a small gradient region with a gradient length of $x_0 <\rho_{s0}$, and assuming that $\left|\nabla_{\perp} \phi_{GK}\right|\sim \frac{T_e}{eN_{e}}\left|\nabla_{\perp} \varrho_i^*\right|\gg  \left|\nabla_{\perp} \phi_{DK}\right|$, we obtain the following condition for an instability,
\begin{align}
    \frac{1}{T_{eDK}}+\frac{1}{T_i}\left[1-\exp\left(-\frac{T_{iDK}k_\perp^2}{\Omega_i^2}\right)I_{0}\left(\frac{T_{iDK}k_\perp^2}{\Omega_i^2}\right)\right]>\frac{\left|\nabla_{\perp} \varrho_i^*\right|}{eN_{DK}\left|\nabla_{\perp} \phi_{GK}\right|}.\label{eq:lin_inst}
\end{align}
The details of the calculation leading to \eqref{eq:lin_inst} are specified in \appref{app:lin_disp}. For small structures where $v_{Thi} k_\perp\gg \Omega_i$, we expect $\phi_{GK}\sim T_{eDK}\varrho_i^*/N_{DK}$. Indeed, in \eqref{eq:Poisson_GK}, using Eqs. (\ref{eq:Poisson_gk_op}) and (\ref{eq:varrho_i}) $\projpj{}{\partial_\mu \gyroavgadj{\gyroavg{\phi_{GK}}}}\propto \exp\left(-T_{i0}k_{\perp}^2/m_i\Omega_i^2\right)\rightarrow 0 $ when $v_{Thi} k_\perp\gg \Omega_i$. Therefore, \eqref{eq:lin_inst} is in general satisfied for large enough $k_{\perp}$. This instability would allow for a non-linear growth of small-scale structures (due to the gradient size, $x_0$, of $\phi_{GK}$ further decreasing). On the other hand, the $\sim \nu_{i0}\nabla_\perp^2$ term in \eqref{eq:C_iGK} and the $\exp\left(-\frac{T_{i0}k_\perp^2}{\Omega_i^2}\right)$ factor in $\varrho_i^*$ in \eqref{eq:varrho_i} inhibits the buildup of large gradient regions for $\phi_{GK}$. As a direct consequence, in the global simulations, this instability is only observed in the simulation with low GK collisionality.



\section{Conclusion}\label{sec:concl}
First of a kind simulations that self-consistently couple full-f drift-kinetic (DK) and $\delta$-f gyrokinetic (GK) fields are performed in a linear plasma device, relying on a gyromoment approach. We consistently derive the ion and electron system of equations starting from the gyro-averaged Boltzmann equations. Adopting a critical balance ordering, we find that the electrons are described by a drift-reduced Braginskii model. The DK and GK parts of the ion distribution function are projected on a Hermite-Laguerre basis yielding a hierarchy equation for the resulting gyromoments of the distribution function.

The physical model is numerically implemented and simulations are carried out using LAPD parameters. Spectral convergence is observed with a small amount of moments, due to the high ion-ion collisionality that leads to an ion distribution function close to a bi-Maxwellian. The inclusion of the small-scale GK part of the electrostatic potential and ion gyromoments does not affect the DK fields at the physical collisionality. When artificially reducing the GK collision term and increasing the GK sources, a small increase in small-scale turbulent structures in the plane perpendicular to the magnetic field is observed for the DK fields.

The GK electrostatic potential is dominated by long-wavelength modes that have significantly smaller perpendicular sizes than its DK counterpart as well as smaller amplitude satisfying the assumptions behind the physical model. The amplitude of the GK gyromoments quickly decreases with increasing order of the Hermite and Laguerre polynomials due to the high ion-ion collisionality.

Linearly, the growth-rates of the system of equations are driven by the DK fields supporting the findings from the global full-f simulations that the GK fields play a subdominant role in driving turbulence. The driving instability has a Kelvin-Helmholz nature. If the amplitude of the GK sources are amplified and collisions for the GK fields are suppressed, increased growth rates are observed caused by the GK density gradients. In this scenario, the growth rate of unstable modes at large perpendicular wavenumbers is amplified. We show analytically that small gradient length scales for the GK potential and density can drive a GK Kelvin-Helmzholz-like instability. Additionally, this instability can non-linearly cause small-scale structures to develop in the full 3D simulations. 

\section*{Acknowledgements}
The authors thank, Louis N. Stenger, Brenno Jason Sanzio Peter De Lucca, Samuel Ernst, and Zeno Tecchiolli for fruitful discussions. Simulations were carried out due to the grant EHPC-REG-2023R03-144 on the Discoverer supercomputer. This work has been carried out within the framework of the EUROfusion Consortium, partially funded by the European Union via the Euratom Research and Training Programme (Grant Agreement No 101052200 — EUROfusion).
The Swiss contribution to this work has been funded in part by the Swiss State Secretariat for Education, Research and Innovation (SERI).
Views and opinions expressed are however those of the author(s) only and do not necessarily reflect those of the European Union, the European Commission or SERI. 
Neither the European Union nor the European Commission nor SERI can be held responsible for them.

\bibliographystyle{jpp}
\bibliography{biblio}


\appendix
\section{Calculation of higher order electron distribution function}\label{app:Fe1_calc}
In order to solve for $F_{eDK1}$, we use the ansatz \eqref{eq:FeDK1_ansatz} with the additional ansatz for $f_{sh}$ in \eqref{eq:fsh_ansatz}. We project \eqref{eq:Boltzmanne1} on the $v_{\|} L_j^{\left(3/2\right)}$ basis, using parity in $v_{\|}$ (notice that the collision operators conserve parity in $\bm v$)
\begin{subequations}
\begin{align}
    \frac{4}{3\sqrt{\pi}}\left(2\Gamma\left(5/2\right)\delta_0^j\left(\nabla_{\|}P_{ e1}+eN_{e1}\nabla_{\|}\phi_{DK}\right)\right.&\nonumber\\
    \left.-2\Gamma\left(7/2\right)\delta_1^j\left(\nabla_{\|}P_{ e1}+e\frac{P_{e1}}{T_{0}}\nabla_{\|}\phi_{DK}\right)\right)&=m_e \left(C_{ee1}^j+C_{ei1}^j\right)\\
    C_{ee1}^j&= v_{Th e}^2\nu_{ee}\sum_{p} K_{ee}^{jp}A_p,\\
    C_{ei1}^j= v_{Th e}^2\nu_{ei}\left(\sum_{p} K_{ei}^{jp}A_p-\frac{m_e}{T_{e0}} K_{ei}^{j0}J_{\|}\right),
    \intertext{and finally,}
    \delta_0^j\left(\nabla_{\|}P_{ e1}+eN_{e1}\nabla_{\|}\phi_{DK}\right)+m_e  \nu_{ei}K_{ei}^{j0}J_{\|}&\nonumber\\
    \frac{5}{2}\delta_1^j\left(\nabla_{\|}P_{ e1}+e\frac{P_{e1}}{T_{e0}}\nabla_{\|}\phi_{DK}\right)= T_{e0} \sum_{k}&\left(\nu_{ee}K_{ee}^{jk}+\nu_{ei}K_{ei}^{jk}\right)A_k,\label{eq:Brag_j}
\end{align}
\end{subequations}
where we have defined \cite{braginskii1965transport,FelixParraBrag}
\begin{subequations}
\begin{align}
&\left(\begin{array}{ccccc}
K^{00}_{e e} & K^{01}_{e e} & K^{02}_{e e} & K^{03}_{e e} & \cdots \\
K^{10}_{e e} & K^{11}_{e e} & K^{12}_{e e} & K^{13}_{e e} & \cdots \\
K^{20}_{e e} & K^{21}_{e e} & K^{22}_{e e} & K^{23}_{e e} & \cdots \\
K^{30}_{e e} & K^{31}_{e e} & K^{32}_{e e} & K^{33}_{e e} & \cdots \\
\vdots & \vdots & \vdots & \vdots & \ddots
\end{array}\right)=\sqrt{2}\left(\begin{array}{ccccc}
0 & 0 & 0 & 0 & \cdots \\
0 & 1 & 3 / 4 & 15 / 32 & \cdots \\
0 & 3 / 4 & 45 / 16 & 309 / 128 & \cdots \\
0 & 15 / 32 & 309 / 128 & 5657 / 1024 & \cdots \\
\vdots & \vdots & \vdots & \vdots & \ddots
\end{array}\right),\\
&\left(\begin{array}{ccccl}
K^{00}_{e i} & K^{01}_{e i} & K^{02}_{e i} & K^{03}_{e i} & \cdots \\
K^{10}_{e i} & K^{11}_{e i} & K^{12}_{e i} & K^{13}_{e i} & \ldots \\
K^{20}_{e i} & K^{21}_{e i} & K^{22}_{e i} & K^{23}_{e i} & \cdots \\
K^{30}_{e i} & K^{31}_{e i} & K^{32}_{e i} & K^{33}_{e i} & \ldots \\
\vdots & \vdots & \vdots & \vdots & \ddots
\end{array}\right)=\left(\begin{array}{ccccc}
1 & 3 / 2 & 15 / 8 & 35 / 16 & \ldots \\
3 / 2 & 13 / 4 & 69 / 16 & 165 / 32 & \ldots \\
15 / 8 & 69 / 16 & 433 / 64 & 1077 / 128 & \ldots \\
35 / 16 & 165 / 32 & 1077 / 128 & 2957 / 256 & \ldots \\
\vdots & \vdots & \vdots & \vdots & \ddots
\end{array}\right).
\end{align}
\end{subequations}
We note that the $j=0$ equation of \eqref{eq:Brag_j} is solved for using $J_{\|}=\mathcal{O}\left(c_{s}n_e\epsilon_{\perp}\right)$ which gives the additional condition by combining \eqref{eq:moment_hierachy_i} with $\left(p,j\right)=\left(1,0\right)$ and \eqref{eq:Brag_j} with $j=0$,
\begin{equation}
    \frac{1}{m_i}\nabla_{\|} P_{\| i DK}-\frac{1}{m_e}\nabla_{\|} P_{e 1}-\frac{eN_{e 1}}{m_e}\nabla_{\|}\phi_{DK}=\nu_{ei} \left(J_{\|}-\frac{T_{e0}}{m_e}\sum_k K_{ei}^{0k}A_k\right)+\mathcal{O}\left(\Omega_i c_s\epsilon\epsilon_{\perp}^2\right).\label{eq:Par_force_balance}
\end{equation}
We obtain from \eqref{eq:Brag_j} the following system,
\begin{align}
    \sum_{k}\left(K_{ee}^{jk}+K_{ei}^{jk}\right)A_k&=\begin{bmatrix}
        C_{P1}\\
        0\\
        0
    \end{bmatrix}+\frac{m_e n_e}{T_e}U_{\| iDK}K_{ei}^{j0},\label{eq:Braginskii_solv}
    \intertext{with}
    C_{P1}&=\frac{1}{T_{e0}\nu_{ee}}\left(\nabla_{\|}P_{ e1}+e\frac{P_{e1}}{T_{e0}}\nabla_{\|}\phi_{DK}\right)=\frac{N_{e0}}{T_{e0}\nu_{ee}}\nabla_{\|}T_{ e1},
\end{align}
where we have defined $T_{e1}=P_{e1}/N_{e0}$ and used \eqref{eq:Ne0_def} to relate $\nabla_{\|} \phi_{DK}$ to $\nabla_{\|} N_{e0}$. The solution to \eqref{eq:Braginskii_solv} truncated at $j=3$ gives \cite{braginskii1965transport,FelixParraBrag}
\begin{subequations}
    \begin{align}
        A_1 &=1.266 C_{P1}+0.286\frac{m_e}{T_{e0}} J_{\|},\\ 
        A_2&= -0.654 C_{P1}+0.055 \frac{m_e}{T_{e0}} J_{\|},\\ 
        A_3&=0.0193 C_{P1}+0.0238 \frac{m_e}{T_{e0}} J_{\|}.
    \end{align}
\end{subequations}
Taking the moments of \eqref{eq:Boltzmanne1} with $1$, and $m_ev^2$ gives

\begin{subequations}
    \begin{align}
    \partial_t N_{e 0}+\frac{1}{B}\poissonbracket{\phi_{DK}+\phi_{GK}}{ N_{e 0}}+\nabla_{\|}\left( N_{e 0} U_{\| iDK}-J_{\|}\right)&=S_N,\label{eq:dt_ne0}\\
    3\partial_t  T_{e1}+\frac{3}{B}\poissonbracket{\phi_{DK}+\phi_{GK}}{ T_{e0}}+\frac{2}{N_{e0}}\nabla_{\|} q_{\| e}&\nonumber\\
    -2U_{\| e DK}\frac{T_{e0}}{N_{e0}}\nabla_{\|}n_{e}-3\frac{T_{e0}}{N_{e0}}\nabla_{\|}\left(N_{e0} U_{\| e DK}\right)&=\C_{ei1}^{mv^2}+2\frac{S_{Pe}}{N_{e0}}-3\frac{T_{e0}}{N_{e0}} S_n,\\
    3\partial_t  T_{e0}+\frac{3}{B}\poissonbracket{\phi_{DK}+\phi_{GK}}{ T_{e0}}+\frac{1}{N_{e0}}\nabla_{\|}\left(2 q_{\| e}+3T_{e0} J_{\|}\right)&\nonumber\\
    -2U_{\| i DK}\frac{T_{e0}}{N_{e0}}\nabla_{\|}N_{e 0}+2J_{\|}\frac{T_{e0}}{N_{e0}^2}\nabla_{\|}N_{e 0}-3\frac{T_{e0}}{N_{e0}}\nabla_{\|}\left(N_{e0} U_{\| i DK}\right)&=\C_{ei1}^{mv^2}+2\frac{S_{Pe}}{N_{e0}}-3\frac{T_{e0}}{N_{e0}} S_n,\label{eq:dt_Te0}
    \intertext{with}
    q_{\| e}=\frac{1}{2}\int d\mathbf{v} m_e v_{\|}v^2 F_{e1}=T_{e0}\left(\frac{5}{2}N_{e 0} U_{\| e DK}-\frac{5}{2}\frac{A_1}{m_e}\right)&,
    \end{align}
\end{subequations}
 $\projpj{}{ S_e}_e=S_{N}$, $\projpj{}{m_ev^2 S_e/2}_e=S_{Pe}$, and $\projpj{}{m_ev^2 C_{ei}/2}_e=\C_{ei1}^{mv^2}$.

Finally, we define $N_{eDK}=N_{e0}+N_{e1}$ and $T_{eDK}=T_{e0}+T_{e1}$, neglecting $\C_{ei1}^{mv^2}$ and adding a few terms to Eqs. (\ref{eq:dt_ne0}) and (\ref{eq:dt_Te0}) noticing that these does not change the order of validity regime of the equations, we arrive at Eqs. (\ref{eq:dt_Ne}) and (\ref{eq:dt_Te}) having defined $S_{T_e}=S_{Pe}/3N_{eDK}-T_{eDK}S_N/N_{eDK}$. It is noted that Eqs. (\ref{eq:dt_ne0}) and (\ref{eq:dt_Te0}) have some higher order terms, however, these are added such that they exactly correspond to the drift-reduced-Braginskii model \cite{FelixParraBrag,Zeiler1997}.

Equation (\ref{eq:Par_force_balance}) represents nothing more than force balance for electrons and ions. In order to maintain quasi neutrality, we need $J_{\|}\sim 0$ which for large $\nu_{ei}$ is obviously satisfied. For low $\nu_{ei}$, if there is an imbalance in the accelerations from the electron and ion pressure, a large electric field builds up to accelerate the electrons such that $U_{\| e}\sim U_{\| i}$.
\section{Hermite-Laguerre projection of gyroaveraged quantities}\label{app:proj_gyro_avg}
For completeness, the complicated expression for the gyroaveraged quantities in Eqs. (\ref{eq:moment_hierachy_gk_i}) and (\ref{eq:Poisson_GK}) are reported. The linear expressions (in $\phi_{GK}$) are given by \cite{Frei2020}
\begin{subequations}
    \begin{align}
        \projpj{}{\partial_\mu \gyroavgadj{\gyroavg{\phi_{GK}}}}_i&=-\sum_{\bm k}\frac{B}{ T_{i0}}\sum_{n\geq 0, m\geq 1}K_{n}\left(b_i\right) K_{m}\left(b_i\right)\nonumber\\
        &\times \sum_{l=0}^{m-1}\sum_{f=0}^{n+l}\bar{d}_{nlf}^0 \N^{0f}_{iDK} \phi_{GK}\left(\bm k\right)e^{i\bm k\cdot\bm x},\label{eq:Poisson_gk_op}\\
        \varrho_i^*&=\sum_{\bm k}\sum_{n\geq 0} K_{n}\left(b_i\right) \N^{0 n}_{iGK}\left(\bm k\right) e^{i\bm k\cdot\bm x},\label{eq:varrho_i}\\
        \projpj{pj}{\gyroavg{\phi_{GK}}}_i&=\sum_{\mathbf{k}}\sum_{n\geq 0}D_{n}^{pjn}\phi_{GK}\left(\mathbf{k}\right)e^{i\bm k\cdot\bm R},\label{eq:projpj_gyroavg}\\
    D_{n}^{pkl}&=\sum_{r=0}^{l+k}\bar{d}_{lkr}^0 \N^{p r}_{iDK} K_{n}\left(b_i\right),\\
    \bar{d}_{r j f}^n&=\sum_{r_1=0}^r \sum_{j_1=0}^j \sum_{f_1=0}^f L_{j j_1}^{-1 / 2} L_{r r_1}^{n-1 / 2} L_{f f_1}^{-1 / 2}\left(r_1+j_1+f_1\right)!,\\
    L_{n l}^m&=\frac{(-1)^l(m+n+1 / 2)!}{(n-l)!(m+l+1 / 2)!l!},\\
    K_n\left(b_i\right)&=\frac{1}{n!}\left(\frac{b_i}{2}\right)^{2n} e^{-\left(b_i/2\right)^2},\quad b_i=\frac{k_{\perp}\sqrt{2 T_{i0}}}{\Omega_i\sqrt{m_i}},
\intertext{while the non-linear projections are}
    \projpj{pj}{\gyroavg{\psi}-\gyroavg{\phi_{GK}}}_i&=\frac{q_i^2}{2m_i\Omega_i}\projpj{pj}{\partial_\mu \left(\left\langle\phi_{GK}\right\rangle_{\mathbf{R}}^2-\left\langle\phi_{GK}^2\right\rangle_{\mathbf{R}}\right)}_i\nonumber\\
    &+\frac{q_i}{2m_i \Omega_i^2}\projpj{pj}{\gyroavg{\left(\mathbf{b}\times\nabla \overline{\widetilde{\phi_{GK}}}\right)\cdot\nabla \widetilde{\phi_{GK}}}}_i,\\
    \projpj{pj}{\partial_\mu \left(\left\langle\phi_{GK}\right\rangle_{\mathbf{R}}^2-\left\langle\phi_{GK}^2\right\rangle_{\mathbf{R}}\right)}_i&=\sum_{\mathbf{k},\mathbf{k}^{\prime}}\sum_{n=0}^{\infty}\frac{1}{n+1}\left[K_{n}\left(\left|\bm b_i+\bm b_i^\prime\right|\right)\left|\bm k_{\perp}+\bm k_{\perp}^\prime\right|^2\sum_{f=0}^{j+n}d^{1}_{njf}\N^{pf}_{iDK}\vphantom{\sum_{m=0}^{\infty}}\right.\nonumber\\
    &\left.-\sum_{m=0}^{\infty}\left(K_{n}\left(b_i\right)K_{m}\left(b_i^{\prime}\right)k_{\perp}^2+K_{n}\left(b_i^{\prime}\right)K_{m}\left(b_i\right)\left(k_{\perp}^{\prime}\right)^2\right)\right.\nonumber\\
    \left.\sum_{f=0}^{m+n}d^{1}_{nmf}\sum_{s=0}^{f+j}d^{0}_{fjs}\N^{ps}_{iDK}\right]&\phi_{GK}\left(\mathbf{k}\right)\phi_{GK}\left(\mathbf{k}^{\prime}\right)e^{i\left(\mathbf{k}+\mathbf{k}^{\prime}\right)\cdot\mathbf{R}},
    \end{align}
\end{subequations}
while the final term is a bit more tricky, without projecting, it can be expressed as

\begin{align}
    \left\langle\left(\mathbf{b}\times\nabla \overline{\widetilde{\phi_{GK}}}\right)\cdot\nabla \widetilde{\phi_{GK}}\right\rangle_{\mathbf{R}}&=2 \sum_{\mathbf{k},\mathbf{k}^{\prime},n> 0}\mathbf{b}\times \mathbf{k}\cdot\mathbf{k}^{\prime} \frac{\left(-1\right)^n}{n}J_n\left(b^{\prime}\right)J_n\left(b\right)\sin n\alpha\nonumber\\
    &\phi_{GK}\left(\mathbf{k}\right)\phi_{GK}\left(\mathbf{k}^{\prime}\right)e^{i\left(\mathbf{k}+\mathbf{k}^{\prime}\right)\cdot\mathbf{R}},\label{eq:Bitch_sum}
\end{align}
where we have introduced $\alpha$ as the angle between $\mathbf{k}$ and $\mathbf{k}^\prime$ such that $\mathbf{k}\cdot\mathbf{k}^\prime=kk^\prime \cos \alpha$ and the Bessel function of the first kind of order $n$, $J_n$. Finally we truncate the sum in \eqref{eq:Bitch_sum} at $n=1$ for simplicity and get
\begin{align}
    \projpj{pj}{\gyroavg{\left(\mathbf{b}\times\nabla \overline{\widetilde{\phi_{GK}}}\right)\cdot\nabla \widetilde{\phi_{GK}}}}_i&=\frac{ T_{i0}}{m_i\Omega_i^2}\left( \mathbf{b}\times \mathbf{k}\cdot\mathbf{k}^{\prime}\right)^2\sum_{n=0}^{\infty}\sum_{m=0}^{\infty}\frac{1}{\left(n+1\right)}K_{n}\left(b_i\right)K_{m}\left(b_i^\prime\right)\nonumber\\
    \times
    \left(\sum_{f=0}^{m+n}d^{1}_{nmf}-\sum_{f=0}^{m+1+n}d^{1}_{n\left(m+1\right)f} \right)&\sum_{s=0}^{f+j}d^{0}_{fjs}\N^{ps}_{iDK}\phi_{GK}\left(\mathbf{k}\right)\phi_{GK}\left(\mathbf{k}^{\prime}\right)e^{i\left(\mathbf{k}+\mathbf{k}^{\prime}\right)\cdot\mathbf{R}}.
\end{align}
Finally for the projection on the GK distribution function, we simply replace $\N^{pj}_{iDK}$ with $\N^{pj}_{iGK}$ in \eqref{eq:projpj_gyroavg}. For all the sums, we truncate maximally at $J_{DK}$ and $J_{GK}$ for the DK and GK projections, respectively \cite{Hoffmann2023,Hoffmann2023b}.

\section{Linear dispersion relation of the GK Kelvin-Helmholz mode}\label{app:lin_disp}
In this appendix, we derive a dispersion relation for the $k_\|=0$, $m\gg 1$ mode observed for the full model in \figref{fig:linear}. We go to a slab geometry with $x$ the radial direction and $y$ the poloidal direction. We assume that the ion DK distribution function is a fixed Maxwellian with density $N_{DK}$ and temperature $T_i$, furthermore, we approximate the ion GK distribution function as an isotropic distribution function, also, with temperature $T_i$. We linearize all fields, $f$, with the equilibrium $\Bar{f}=\Bar{f}\left(x\right)$ and the fluctuating part $\Tilde{f}\left(x\right)\exp\left(-i\omega t+i k_y y\right)$. We use \eqref{eq:dt_Omega}, perform the gyroaverage operation on \eqref{eq:moment_hierachy_gk_i} with $\left(p,j\right)=\left(0,0\right)$ while only including terms linear in $\epsilon_\delta$. Finally, we close the system with \eqref{eq:Poisson_GK} and to get
\begin{subequations}
\begin{align}
    &\nabla_{\perp}\cdot\left(-i\omega^*\Bar{N_{DK}}\nabla_{\perp}\Tilde{\phi_{DK}} -ik_y^* \Bar{N_{DK}}\partial_x\nabla_{\perp}\Bar{\phi_{DK}}\Tilde{\phi_{tot}} \right)=0,\label{eq:lin_vorticity}\\
      &-i\omega \Tilde{\varrho_{i}^*}+ik_y^*\left(\partial_x\Bar{\phi_{DK}}\Tilde{\varrho_{i}^*}-\partial_x \Bar{\varrho_{i}^*}\Tilde{\phi_{DK}}\right)=0,\\
      &\Tilde{\varrho_{i}^*}=\Bar{N_{DK}}I_y\Tilde{\phi_{GK}},\\
\intertext{with}
    &I_y=e\left(\frac{1}{T_{e}}+\frac{1}{T_{i}}-\frac{1}{T_{i}}\exp\left(-\frac{b_{iy}^2}{2}\right)I_{0}\left(\frac{b_{iy}^2}{2}\right)\right),\\
    k_y^*&=\frac{k_y}{B\rho_*},\quad \phi_{tot}=\phi_{DK}+\phi_{GK},\quad \text{and} \quad \omega^*=\omega-k_y^*\partial_x\Bar{\phi_{tot}}.
\end{align}
\end{subequations}
Neglecting the long-wavelength contribution of \eqref{eq:lin_vorticity}, i.e. only keeping terms linear in $\phi_{GK}$ and combining the equations lead to
\begin{align}
    -\nabla_\perp\cdot\left(k_y^*\partial_x\Bar{\phi_{GK}}\nabla_\perp \Tilde{\phi_{DK}}+\frac{\left(k_y^*\right)^2}{\left(\omega-k_y^*\partial_x\Bar{\phi_{DK}}\right)I_y}\partial_x\nabla_{\perp}\Bar{\phi_{DK}}\partial_x \Bar{\varrho_{i}^*}\Tilde{\phi_{DK}}\right)=0.\label{eq:C2}
\end{align}
Assuming that $\partial_x\phi_{DK}$ and $\partial_x\phi_{GK}$ are piecewise linear around a region of $x_0$
\begin{subequations}
\begin{align}
    \partial_x\Bar{\phi_{DK }}&=\begin{cases}
        -V_{0},\quad &x<-x_0,\\
        V_0\frac{x}{x_0},\quad &-x_0<x<x_0,\\
         V_0,\quad & x>x_0,
         \end{cases},\\
    \partial_x\Bar{\phi_{GK }}&=\begin{cases}
        -V_{1},\quad &x<-x_0,\\
        V_1\frac{x}{x_0},\quad &-x_0<x<x_0,\\
         V_1,\quad & x>x_0,
         \end{cases},
\end{align}
\end{subequations}
while $\Tilde{\phi_{DK}}=V_n$ is constant. We assume that $V_0\ll V_1$ such that a general solution for $\Tilde{\phi_{DK}}$ is
\begin{align}
    &\Tilde{\phi_{DK}}=\begin{cases}
        \left(A_1+A_2\exp\left(2k_yx_0\right)\right)\exp\left(k_y x\right),\quad &x<-x_0,\\
        A_{1}\exp\left(k_y x\right)+A_{2}\exp\left(-k_y x\right),\quad &-x_0<x<x_0,\\
         \left(A_2+A_1\exp\left(2k_yx_0\right)\right)\exp\left(-k_y x \right),\quad & x>x_0,
    \end{cases}
\end{align}
where we have used that $\Tilde{\phi_{DK}}$ vanishes at $-\infty$ and $\infty$ and is continuous. Finally, we integrate \eqref{eq:C2} over each discontinuity ($x=\pm x_0$) and obtain a matching condition
\begin{align}
    \begin{bmatrix}
        \frac{\left(k_y^*\right)^2V_0V_n}{\left(\omega+k_y^* V_0\right)x_0 I_y}e^{-2 k_y x_0}& -2k_y k_y^*V_1+\frac{\left(k_y^*\right)^2V_0V_n}{\left(\omega+k_y^* V_0\right)x_0 I_y}\\
        2k_y k_y^*V_1+\frac{\left(k_y^*\right)^2V_0V_n}{\left(\omega-k_y^* V_0\right)x_0 I_y}& \frac{\left(k_y^*\right)^2V_0V_n}{\left(\omega-k_y^* V_0\right)x_0 I_y}e^{-2 k_y x_0}
    \end{bmatrix}\cdot\begin{bmatrix}
        A_1\\
        A_2
    \end{bmatrix}=0.\label{eq:Matrix_prob}
\end{align}

For $\Tilde{\phi_{DK}}$ to be able to grow, this system of equations needs to have infinitely many solutions, implying that the determinant of the matrix in \eqref{eq:Matrix_prob} needs to be 0. This requirement restricts $\omega$
\begin{align}
    \omega^2&=\left(\frac{k_y^* V_0V_n}{k_y V_1 x_0 I_y}-2k_y^* V_0\right)^2-\left(\frac{k_y^* V_0V_n}{k_y V_1 x_0 I_y}\right)^2 e^{-4 k_y x_0}.
\end{align}
For the instability to grow, we need $\omega^2<0$ which implies
\begin{align}
    \left(\frac{k_y^* V_0V_n}{k_y V_1 x_0 I_y}-2k_y^* V_0\right)^2<\left(\frac{k_y^* V_0V_n}{k_y V_1 x_0 I_y}\right)^2 e^{-4 k_y x_0}.
\end{align}
As can be seen, if $V_n$ and $V_1$ have the same sign, this relation can only be satisfied for small $k_y x_0<1/4$ (necessary but not sufficient condition). In the limit $x_0\rightarrow 0$, we get $1>V_n/V_1 I_y$ and the instability condition in \eqref{eq:lin_inst} is recovered. It is noted that this instability is limited by diffusion (numeric and physical through the collision operator) which restricts the development of small $x_0$. Finally, the instability gives rise to more small-scale gradients in $\Tilde{\phi_{DK}}$ which can drive a cascade of small-scale instabilities.

\end{document}